\documentclass[10pt,aps,prl,twocolumn,superscriptaddress,preprintnumbers]{revtex4-1}

\usepackage{amsmath,amssymb,amsfonts}
\usepackage{graphicx,color}
\usepackage{verbatim}
\usepackage{float}
\usepackage{dcolumn}
\usepackage{natbib}
\usepackage{bm}
\usepackage{hyperref}
\usepackage[table]{xcolor}

\definecolor{Gray}{gray}{0.85}
\newcommand{\ie}{\emph{i.e.}}

\newcommand{\EC}{{\em E. coli}}

\begin{document}

\title{Essentiality, conservation, evolutionary pressure and codon bias in bacterial genes}

\author{Maddalena Dilucca}
\affiliation{Dipartimento di Fisica, Sapienza University of Rome, Rome -- Italy}
\email{maddalena.dilucca@roma1.infn.it}
\author{Giulio Cimini}
\affiliation{IMT School for Advanced Studies, Lucca -- Italy}
\affiliation{Istituto dei Sistemi Complessi (ISC)-CNR, Rome -- Italy}
\author{Andrea Giansanti}
\affiliation{Dipartimento di Fisica, Sapienza University of Rome, Rome -- Italy}
\affiliation{INFN Roma1 unit, Rome -- Italy}

\begin{abstract}
Essential genes constitute the core of genes which cannot be mutated too much nor lost along the evolutionary history of a species. 
Natural selection is expected to be stricter on essential genes and on conserved (highly shared) genes, than on genes that are either nonessential or peculiar to a single or a few species. 
In order to further assess this expectation, we study here how essentiality of a gene is connected with its degree of conservation among several unrelated bacterial species, 
each one characterised by its own codon usage bias. Confirming previous results on \EC, 
we show the existence of a universal exponential relation between gene essentiality and conservation in bacteria. 
Moreover we show that, within each bacterial genome, there are at least two groups of functionally distinct genes,  
characterised by different levels of conservation and codon bias: i) a core of essential genes, mainly related to cellular information processing; 
ii) a set of less conserved nonessential genes with prevalent functions related to metabolism. 
In particular, the genes in the first group are more retained among species, are subject to a stronger purifying conservative selection and display a more limited repertoire of synonymous codons. 
The core of essential genes is close to the minimal bacterial genome, which is in the focus of recent studies in synthetic biology, 
though we confirm that orthologs of genes that are essential in one species are not necessarily essential in other species. 
We also list a set of highly shared genes which, reasonably, could constitute a reservoir of targets for new anti-microbial drugs.
\keywords{Bacteria \and Essentiality \and Conservation \and Selective pressure \and Codon bias}
\end{abstract}

\maketitle 


\section*{Introduction}

From an evolutionary point of view, all living species are in a process of adaptation to the environments they happen to live in. 
This process rests on the incorporation of genetic mutations into the genomes at the level of populations of species, 
which evolves on time-scales far longer than the time-scale of a generation. 
Signals from this process can be searched for in the sequences of single genes, of several genes within one single species, and among several species. 
In a previous work we have shown that, in \EC, essentiality and degree of conservation of genes are subtly correlated with the codon bias displayed by their sequences \citep{Dilucca2015}. 
In this work we extend those observations to a set of unrelated bacterial species, by elaborating on the connection between gene essentiality and conservation, and their relation with codon bias.

Individual genes in the genome of a given species contribute differentially to the survival and propagation of the organisms of that species. 
According to their known functional profiles and based on experimental evidences, genes can be divided into two categories: essential and nonessential ones \citep{Gerdes2003,Fang2005}. 
Essential genes are not dispensable for the survival of an organism in the environment it lives in and the functions they encode are, 
therefore, considered as fundamental for life, irrespective of environmental changes \citep{Fang2005,Peng2014}. 
On the other hand, nonessential genes are those which are dispensable \citep{Lin2010}, being related to functions that can be silenced without lethal effects for the phenotype. 
Naturally, each species has adapted to one or more evolving environments and, plausibly, genes that are essential for one species may be not essential for another one. 
However, the set of genes that are essential in several bacterial species should encode for functions that are fundamental for life.
As suggested by a quite broad literature, essential genes are more conserved than nonessential ones \citep{Hurst1999,Jordan2002,Luo2015,Ish-Am2015,Alvarez-Ponce2016}. 
It is worth noting that the term ``conservation'' has at least a twofold meaning. On the one hand, a gene is conserved if orthologous copies are found in the genomes 
of many species, as measured by the Evolutionary Retention Index (ERI) \citep{Gerdes2003,Bergmiller2012}. On the other hand, a gene is (evolutionarily) 
conserved when it is subject to a purifying evolutionary pressure which disfavors mutations \citep{Hurst1999,Hurst2002}. In this second meaning a conserved gene is, generically, a slowly evolving gene.
The ratio $K_a/K_s$ of the number of nonsynonymous substitutions per nonsynonymous site to the number of synonymous substitutions per synonymous site is a widely used measure, 
though not exempt from criticism, to assess whether a gene is under Darwinian selection or a purifying evolutionary pressure. 
 
Beyond essentiality and conservation, in our analysis we also consider the degeneracy of the genetic code, due to the fact that the same amino acid is encoded by different codon triplets (synonymous codons). 
Usage frequencies of synonymous codons vary significantly between different organisms, and also between proteins within the same organism \citep{Kanaya2001}.
This phenomenon, known as {\it codon usage bias}, can be measured by various indices (see \cite {Roth2012} for an overview); 
we use here statistical indicators such as the effective number of codons and the relative synonymous codon usage. 

Our analysis reveals that those genes which are more conserved among bacterial species are also prone to be essential. 
Moreover, the codon usage in these conserved genes is, in general, more optimized than in less conserved genes. 
We have also shown that essential, conserved genes tend to be subject to a relatively more purifying evolutionary pressure. 
We argue that the set of genes with the highest degree of conservation (\ie, those with $\mbox{ERI}=1$) could include putative novel targets for novel anti bacterial strategies, 
as suggested with rather similar arguments by \citet{Dotsch2010}.

\begin{table*}[htbp]
\centering
\caption{{\bf List of bacterial genomes.} For each genome we report organism name, abbreviation, class, ncBI RefSeq, size (number of genes) and percentage of COG. 
Classes are: \emph{Alphaproteobacteria} (1), \emph{Betaproteobacteria} (2), \emph{Gammaproteobacteria} (3), \emph{Epsilonproteobacteria} (4), \emph{Actinobacteria} (5), \emph{Bacilli} (6), 
\emph{Bacteroidetes} (7), \emph{Clostridia} (8), \emph{Deinococci} (9), \emph{Mollicutes} (10), \emph{Spirochaetales} (11), \emph{Aquificae} (12), \emph{Cyanobacteria} (13), \emph{Chlamydiae} (14), 
\emph{Fusobacteria} (15), \emph{Thermotoga} (16). Asterisks denote genomes considered in~\cite{Gerdes2003}. For those genomes annotated in the Database of Essential Genes (DEG, 
highlighted in gray) we report the number of essential (E) and nonessential (NE) genes, as well as the coverage of essentiality. }
\small
\vspace{1cm}
\begin{tabular}{|l|l|r|l|r|r|r|r|r|} 
\hline						
{\bf Organism}	&	{\bf abbr.} & {\bf class} &{\bf ncBI RefSeq}  &{\bf size}  & {\bf \% COG} & {\bf E} &{\bf NE} &{\bf cov. (\%)} \\
\hline\hline															
Agrobacterium tumefaciens (fabrum)	&	agtu*	&	1	&	 nc\_003062 	&	2765	&	83.34  &	&		&		\\
Aquifex aeolicus VF5	&	aqae*	&	12	&	 nc\_000918	&	1497	&	86.65 &		&		&		\\
\rowcolor{Gray}															
Bacillus subtilis 168	&	basu* 	&	6	&	 nc\_000964	&	4175 &	76.84	&	271	&	3904	&	100	\\
\rowcolor{Gray}															
Bacteroides thetaiotaomicron VPI-5482	&	bath 	&	7	&	nc\_004663	&	4778	&	68.22 &	325	&	4453	&	100	\\
Brucella melitensis bv. 1 str. 16M	&	brme*	&	1	&	 nc\_003317.1	&	2059	&	93.50 &		&		&		\\
Buchnera aphidicola Sg uid57913	&	busg*	&	3	&	nc\_004061	&	546	&	100	&		&		&		\\
\rowcolor{Gray}															
Burkholderia pseudomallei K96243	&	bups 	&	2	&	 nc\_006350	&	3398 &	88.80	&	423	&	2932	&	98.8	\\
\rowcolor{Gray}															
Burkholderia thailandensis E264	&	buth  	&	2	&	 nc\_007651	&	3276 &	81.76	&	364	&	2912	&	100	\\
\rowcolor{Gray}															
Campylobacter jejuni 	&	caje* 	&	4	&	 nc\_002163	&	1572	&	82.49 &	222	&	1350	&	100	\\
\rowcolor{Gray}															
Caulobacter crescentus	&	cacr*	&	1	&	 nc\_011916	&	3885	&	65.55 &	402	&	2649	&	78.5	\\
Chlamydia trachomatis D/UW-3/CX	&	chtr*	&	14	&	nc\_000117.1	&	894	&	71.75 &		&		&		\\
Clostridium acetobutylicum ATCC 824	&	clac*	&	8	&	 nc\_003030.1	&	3602	&	77.80 &		&		&		\\
Corynebacterium glutamicum ATCC 13032	&	cogl*	&	5	&	nc\_003450.3	&	2959 &	74.54	&		&		&		\\
Deinococcus radiodurans R1	&	dera*	&	9	&	 nc\_001263.1	&	2629	 &	72.86 &		&		&		\\
\rowcolor{Gray}															
Escherichia Coli K-12 MG1655	&	esco*	&	3	&	 nc\_000913.3	&	4004	 &	86.98 &	587	&	2907	&	87.3	\\
\rowcolor{Gray}															
Francisella novicida U112	&	frno 	&	3	&	 nc\_008601	&	1719 &	82.71 &	390	&	1329	&	100	\\
Fusobacterium nucleatum ATCC 25586	&	funu*	&	15	&	 nc\_003454.1	&	1983	&	79.65 &		&		&		\\
\rowcolor{Gray}															
Haemophilus influenzae Rd KW20	&	hain* 	&	3	&	 nc\_000907.1	&	1610	&	93.28 &	625	&	503	&	70	\\
\rowcolor{Gray}															
Helicobacter pylori 26695	&	hepy* 	&	4	&	 Nc\_000915.2	&	1469	&	76.90 &	305	&	1065	&	93.3	\\
isteria monocytogenes EGD-e	&	limo*	&	6	&	nc\_003210.1	&	2867	&	84.33 &		&		&		\\
Mesorhizobium loti MAFF303099	&	melo*	&	1	&	nc\_002678.2	&	6743	 &	80.33 &		&		&		\\
\rowcolor{Gray}															
Mycoplasma genitalium G37	&	myge 	&	10	&	 nc\_000908	&	475	&	80.84 &	378	&	94	&	99.37	\\
Mycoplasma pneumoniae M129	&	mypn*	&	10	&	nc\_000912.1	&	648	 &	68.62 &		&		&		\\
\rowcolor{Gray}															
Mycoplasma pulmonis UAB CTIP	&	mypu 	&	10	&	nc\_002771	&	782	&	71.57 &	309	&	321	&	80.56	\\
\rowcolor{Gray}															
Mycobacterium tuberculosis H37Rv	&	mytu*	&	5	&	nc\_000962.3	&	3936	&	74 &	592	&	2892	&	88.5	\\
Neisseria gonorrhoeae FA 1090 uid57611	&	nego*	&	2	&	nc\_002946	&	1894 &	76.07	&		&		&		\\
\rowcolor{Gray}															
Porphyromonas gingivalis ATCC 33277	&	pogi	&	7	&	 nc\_010729	&	2089	&	65.46 &	463	&	1626	&	100	\\
\rowcolor{Gray}															
Pseudomonas aeruginosa UCBPP-PA14	&	psae* 	&	3	&	 nc\_008463	&	5892	 &	82.97 &	335	&	4461	&	81.4	\\
Ralstonia solanacearum GMI1000	&	raso*	&	2	&	 nc\_003295.1	&	3436 &	81.22	&		&		&		\\
Rickettsia prowazekii str. Madrid E	&	ripr*	&	1	&	nc\_000963.1	&	8433	&	87.76 &		&		&		\\
\rowcolor{Gray}															
Salmonella enterica serovar Typhi	&	saen 	&	3	&	 nc\_004631	&	4352	 &	78.28 &	358	&	3992	&	99.96	\\
\rowcolor{Gray}															
Shewanella oneidensis MR-1	&	shon 	&	3	&	 nc\_004347	&	4065 &	69.68	&	402	&	1032	&	32.28	\\
Sinorhizobium meliloti 1021	&	sime*	&	1	&	 nc\_003047.1	&	3359	 &	90.26 &		&		&		\\
\rowcolor{Gray}															
Sphingomonas wittichii RW1	&	spwi 	&	1	&	nc\_009511	&	4850	 &	83.89 &	535	&	4315	&	100	\\
\rowcolor{Gray}															
Staphylococcus aureus N315	&	stau* 	&	6	&	nc\_002745.2	&	2582	 &	81 &	302	&	2280	&	100	\\
\rowcolor{Gray}															
Staphylococcus aureus ncTC 8325	&	stau\_ 	&	6	&	nc\_007795	&	2767 &	71.25	&	345	&	2406	&	100	\\
Streptococcus pneumoniae TIGR4	&	stpn*	&	9	&	 nc\_003028.3	&	1814	&	85 &		&		&		\\
\rowcolor{Gray}															
Streptococcus pyogenes MGAS5448	&	stpy 	&	6	&	 nc\_007297	&	1865	 &	77.52 &	227	&	1337	&	83.86	\\
\rowcolor{Gray}															
Streptococcus pyogenes NZ131	&	stpy\_	&	6	&	 nc\_011375	&	1700	&	80.45 &	241	&	1177	&	83.41	\\
\rowcolor{Gray}															
Streptococcus sanguinis	&	stsa 	&	6	&	 nc\_009009	&	2270	&	79.94 &	218	&	2052	&	100	\\
Synechocystis sp. PCC 6803	&	sysp*	&	13	&	nc\_000911.1	&	3179	&	76.96 &		&		&		\\
Thermotoga maritima MSB8	&	thma*	&	16	&	nc\_000853.1	&	1858 &	86.64	&		&		&		\\
Treponema pallidum Nichols	&	trpa*	&	11	&	nc\_000919.1	&	1036	&	71.50 &		&		&		\\
\rowcolor{Gray}															
Vibrio cholerae N16961	&	vich* 	&	3	&	nc\_002505	&	2534	&	85 &	447	&	2079	&	99.68	\\
Xylella fastidiosa 9a5c	&	xyfa*	&	3	&	nc\_002488	&	2766	 &	62.96 &		&		&		\\
\hline															
\end{tabular}
\label{tab.dataset}
\end{table*}

\section{Materials and Methods}\label{sec:1}

\subsection{Bacterial genomes}\label{sec:2}

In this work we consider a set of 45 bacterial genomes from unrelated species, whose details are provided in Table~\ref{tab.dataset}. 
Nucleotide sequences from complete bacterial genomes were downloaded from the FTP server of the National Center for Biotechnology 
Information (NCBI) (\url{ftp://ftp.ncbi.nlm.nih.gov/genomes/archive/old_genbank/Bacteria/}) \citep{Benson2012}.
Note that 31 of the 45 species we collected are also present in the dataset selected by \citet{Gerdes2003} in their seminal paper on \EC's essential genes.

\subsection{Conservation and essentiality}\label{sec:3}

We use the Evolutionary Retention Index (ERI) of \citet{Gerdes2003} as a proxy for gene conservation. 
We compute the ERI of a gene as the fraction of genomes in Table~\ref{tab.dataset} that have at least an ortholog of the given gene. 
A low ERI value means that a gene is specific, common to a small number of genomes, whereas, high ERI is a characteristic of highly shared, conserved, possibly universal genes.

In order to investigate gene essentiality we use the Database of Essential Genes (DEG), available at \url{www.essentialgene.org} \citep{Luo2015}. 
DEG classifies a gene as either essential or nonessential on the basis of a combination of experimental evidence (null mutations or trasposons) and general functional considerations. 
DEG collects genomes from Bacteria, Archea and Eukarya, with different degrees of coverage \citep{Zhang2009,Zhang2014}. 
Of the 45 bacterial genomes we have collected, only 24 are covered---in toto or partially---by DEG, as indicated in Table~\ref{tab.dataset}. 

\subsection{Clusters of orthologous genes}\label{sec:4}

We use the database of orthologous groups of proteins (COGs), available at \url{http://ncbi.nlm.nih.gov/COG/}, for the functional annotation of gene sequences \citep{Galperin2015}. 
We consider 15 functional classes given by COGs, excluding the generic categories R and S for which functional annotation is too general or missing. 
Given a group of genes in a genome, we evaluate the conditional probability that these genes belong to a specific COG as:
\begin{equation}\label{eq:ante-chi}
P(\mbox{COG}|\mbox{group})=P(\mbox{group}|\mbox{COG})P(\mbox{COG})/P(\mbox{group}),
\end{equation}
where $P(\mbox{group})$ is the size of the group with respect to the genome, $P(\mbox{COG})$ is the fraction of the genome belonging to the COG, 
and $P(\mbox{group}|\mbox{COG})$ is the fraction of genes in a given COG that belong to the group.

\subsection{K$_a$/K$_s$}\label{sec:5}

$K_a$/$K_s$ is the ratio of nonsynonymous substitutions per nonsynonymous site ($K_a$) to the number of synonymous substitutions per synonymous site ($K_s$) \citep{Hurst2002}. 
This parameter is widely accepted as a straightforward and effective way of separating genes subject to purifying selection ($K_a/K_s<1$) 
from genes subject to positive Darwinian selection ($K_a/K_s>1$). There are different methods to evaluate this ratio, though the alternative approaches are quite consistent among themselves. 
For the sake of comparison, we have used here the $K_a/K_s$ estimates by \citet{Luo2015} which are based on method by \citet{Nei1986}. 
Note that each genome has a specific average level of $K_a$/$K_s$. 

To study the patterns of $K_a$/$K_s$ in a given group of genes, we use Z-score values:
\begin{equation}
Z_g[(K_a/K_s)|\mbox{group}]=\frac{\langle K_a/K_s\rangle_{\mbox{\footnotesize group},g}-\langle K_a/K_s\rangle_g }{ \sigma_g/ \sqrt{N_g}},
\end{equation}
where $\langle K_a/K_s\rangle_{\mbox{\footnotesize group},g}$ is the average of the ratio within a given group in a genome $g$, 
$\langle K_a/K_s\rangle_g$ and $\sigma_g$ are the average value of $K_a/K_s$ and its standard deviation over the whole genome $g$, 
and $N_g$ is the number of genes in the genome (we use the standard deviation of the mean as we are comparing average values). 
In the following we will analyze $K_a$/$K_s$ patterns with respect to group of genes with similar values of ERI, as indicated by $Z_g[(K_a/K_s)|\mbox{ERI}]$ 
(eventually discerning essential and nonessential genes), as well as with respect to COGs, as indicated by $Z_g[(K_a/K_s)|\mbox{COG}]$. 

Note that $K_a$/$K_s$ values can differ much by magnitude not only among genes in different genomes but even for genes in the same genome, 
and their genome-specific distribution is rather broad with a high peak at zero (see below). 
In such a situation, larger values can in principle bias arithmetic averages. On the other hand, using alternative methods like the geometric mean, the arithmetic mean of the logarithms 
or the harmonic mean cannot be used because of the frequent zero values. Using the median and the median absolute deviation in a Mann-Whitney-Wilcoxon U test 
instead of arithmetic means and standard deviations in Z-score tests to compare distributions leads to results which are highly consistent 
with what we obtained with Z-scores, though more noisy due to the vanishing of the median in many distributions.

\subsection{Codon bias}\label{sec:6}

There are several methods and indices to estimate the degree of codon usage bias in a gene. For an overview of current methods, their classification and rationale see \citet{Roth2012}. 
We use here two basic statistical indicators: the Number of Effective Codons ($N_c$) and the Relative Synonymous Codon Usage ($RSCU$).

$N_c$ measures of the effective diversity of the codons used to code a given 
protein \citep{Wright1990}. In principle, $N_c$ ranges from 20 (when just one single codon is used to code each one of the amino acids) 
to 61 (when the entire degeneracy of the genetic code is fully deployed, and each amino acid is coded by all its synonymous codons on an equal footing). 
Given a sequence of interest, the computation of $N_c$ starts from $F_{\alpha}$, a quantity defined for each family $\alpha$ of synonymous codons (one for each amino acid):
\begin{equation}
F_{\alpha}=\sum_{k=1}^{m_{\alpha}} \left ( \frac{n_{k \alpha}}{n_{\alpha}} \right )^2,
\label{eq:F_cf}
\end{equation}
where $m_\alpha$ is the number of different codons in $\alpha$ (each one appearing $n_{1_\alpha},n_{2_\alpha},\dots,n_{m_\alpha}$ times in the sequence) 
and $n_\alpha=\sum^{m_\alpha}_{k=1}n_{k_\alpha}$. $N_c$ then weights these quantities on a sequence:
\begin{equation}
N_c=N_s+ \frac{K_2 \sum_{\alpha =1}^{K_2} n_\alpha} {\sum_{\alpha =1}^ {K_2} \left (n_{\alpha} F_{\alpha} \right )} 
+ \frac{K_3 \sum_{\alpha =1}^{K_3} n_\alpha} {\sum_{\alpha =1}^ {K_3} \left (n_{\alpha} F_{\alpha} \right )} 
+ \frac{K_4 \sum_{\alpha =1}^{K_4} n_\alpha} {\sum_{\alpha =1}^ {K_4} \left (n_{\alpha} F_{\alpha} \right )} 
\end{equation}
where $N_s$ is the number of families with one codon only and $K_m$ is the number of families with degeneracy $m$ 
(the set of 6 synonymous codons for leucine can be split into one family with degeneracy 2, similar to that of phenylalanine, and one family with degeneracy 4, similar to that of proline). 
In this paper we evaluate $N_c$ by using the implementation provided in DAMBE 5.0 \citep{Xia2013}. 

The relative synonymous codon usage ($RSCU_i$) of each codon $i$ is estimated as:
\begin{equation}
RSCU_i= \frac{X_i}{\frac{1}{N_i} \sum_{j=1}^{n_i} X_j}
\end{equation}
where $X_i$ is the number of occurrences, either in a gene or in the whole genome, of codon $i$. The sum in the denominator runs over $n_i$, the degeneracy of the family of synonymous codons $i$ belongs to. 
For each codon $i$, its $RSCU_i$ is comprised between zero (no usage) and 1 (when only that codon is used among its synonymous alternatives). 
We evaluate these values by using DAMBE 5.0 \citep{Xia2013}. 

The $RSCU$ values of the various codons can be grouped together as the 64 components (including the start codon ATG and the stop codons TAA, TAG and TGA---which are differently used by different species) 
of vectors which measure codon usage bias in a given bacterial species.
To detect different patterns of codon usage between species we use heat maps drawn with CIMMiner (\url{http://discover.nci.nih.gov/cimminer}), 
and we cluster $RSCU$ vectors using Euclidean distances and the Average Linkage cluster algorithm.

\section*{Results and Discussion}

\subsection{Essentiality and conservation in bacterial genes}
\begin{figure}
   \includegraphics[width=8.6cm]{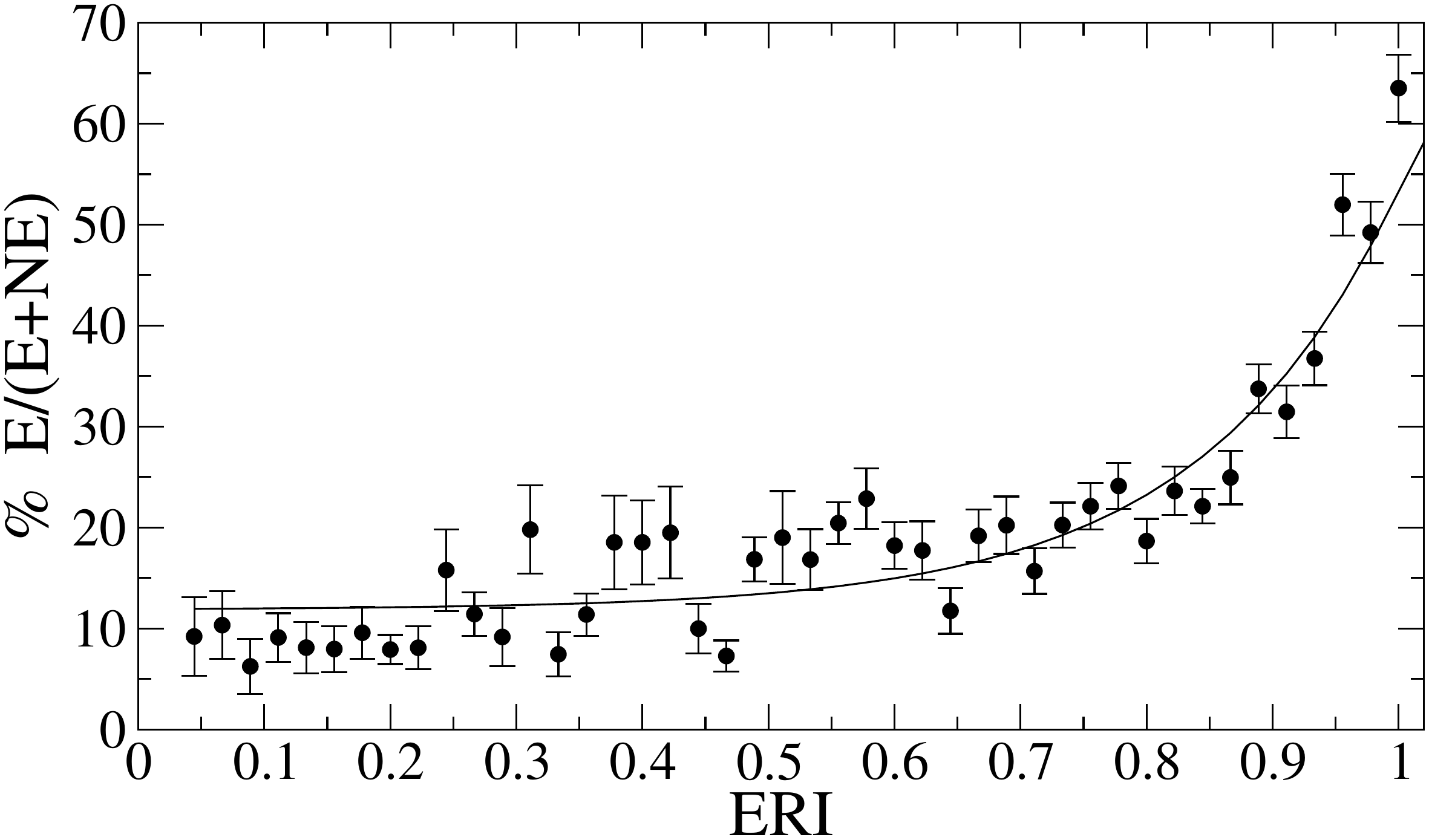}
\caption{Degree of essentiality versus ERI values.
Genes from the 24 DEG-annotated genomes in Table~\ref{tab.dataset} are aggregated 
into bins of ERI, and the fraction of essential genes in each bin is computed. 
Error bars are root mean square deviations, expressing the 
variability in percent essentiality from genome to genome. The solid line shows 
the exponential fit of the data $y=y_0+A\exp(Bx)$, which returns 
$y_0=11.(8)$, $A=0.06(3)$, $B=6.4(8)$ and a coefficient of determination $R^2=0.90$. 
Note that since in small genomes essential genes 
outnumber nonessential ones (as we show below), we exclude from fitted data 
the two small genomes of $myge$ and $mypu$. We also exclude $stpn$ because it is 
poorly covered by DEG.}
\label{fig1}       
\end{figure}

Figure~\ref{fig1} shows the percentage of essential genes within genes with a given value of ERI (which we recall operationally encodes the degree of conservation of a gene). 
The observed exponential dependence generalises to several unrelated species a basic result on \EC, by \citet{Gerdes2003} (see Figure~3 therein), 
and the fit parameters we find are strictly consistent with those reported in that paper. 
This points to the existence of a universal exponential correlation between gene essentiality and conservation in bacteria. 
Indeed, the fact that essential genes should be more evolutionarily conserved than nonessential ones has been previously shown, following different approaches \citep{Jordan2002,Gong2008}. 
Our result confirms those earlier observations and leads to conclude that the more a gene is shared, the more it is likely to be essential. 
This point will be further investigated in the next section.

\begin{figure}
  \includegraphics[width=8.6cm]{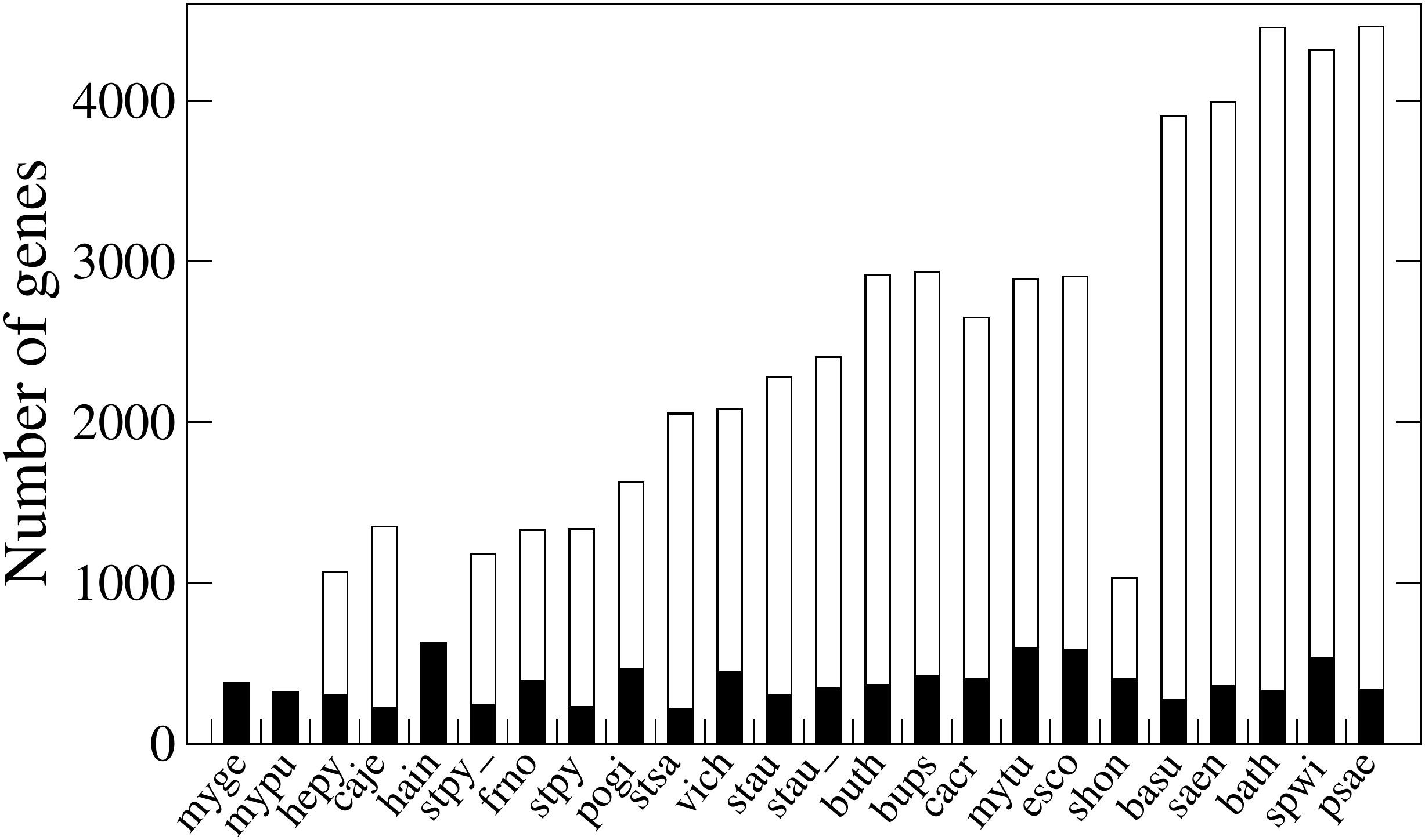}
\caption{Essential and nonessential genes in each bacteria. On the horizontal axis, DEG-annotated species from Table~\ref{tab.dataset} are sorted 
according to the size of their genome. Black and white bars represent the number of essential 
and nonessential genes. The number of essential genes is basically constant in all 
species (average value = 378$\pm$115), while the number of nonessential genes increases with the size of the genomes.}
\label{fig2}      
\end{figure}

Figure~\ref{fig2} further shows that the number of essential genes is rather constant among bacterial genomes. 
In small genomes with less than 1000 genes, most genes are essential. Then, as the size of the genome increases, the number of nonessential genes increases proportionally. 
Note that $shon$ does not follow the trend: it is a species with a peculiar metabolism and, at present, is poorly covered by DEG. Independently from the genome size, 
each bacterial species has a core of about 400 essential genes. 
This observation can be related to recent experiments in synthetic biology, devoted to the in vitro assembly of artificial bacteria with minimal genomes, 
limited to those genes which are necessary to sustain basic life processes \citep{Gil2004,Glass2006,Hutchison2016}. 
In particular, the synthetic bacterium designed and synthesized by \citet{Hutchison2016} has a genome constituted by 473 genes from {\it Mycoplasma mycoides}, 
a species whose genome contains 475 genes and which is evolutionarily close to the {\it Mycoplasma genitalium} considered here ($myge$). 
Of the genes in myge, 80\% are annotated as essential and the remaining 20\% have no annotation yet; 
clearly there are still unknown functions that could be, nevertheless, essential for life.

It is tempting to suppose that the core of essential genes in the bacterial species of Figure~\ref{fig2} 
constitutes a kind of minimal, universal and conserved genome, made by genes that have an orthologous in all species. 
But this is not the case. We have checked that only 83 genes are strictly retained 
($\mbox{ERI}=1$) among all the DEG-annotated bacterial species we consider (and are reported in Table \ref{tab.core}). 
Among them, no one is essential in all species, but only in a fraction $f(\mbox{E})$ of the bacteria. 
Thus, essentiality does not  strictly imply orthology: genes that are essential for one species may be not essential for another one. 
Indeed, the experimentally determined essentiality might disagree between species for a variety of reasons, 
for instance because experimental conditions that are near-optimal for one species are instead demanding for another species.
Concerning strictly retained genes in general, as shown in Table~\ref{tab.func} they have a quite restricted repertoire of functions, 
limited to COGs J ({\it translation, ribosomal structure and biogenesis}: 49 cases), K ({\it transcription}: 7 cases), 
L ({\it replication, recombination and repair}: 7 cases) and O ({\it Post translational modification, protein turnover, chaperones}: 8 cases). 
Hence, more than half of these genes correspond to ribosomal proteins with different degree of shared essentiality, as evaluated by $f(\mbox{E})$. 
We have checked in DrugBank that several of these genes, as expected, are targets of antimicrobial drugs in \EC, 
as shown with bold COG ids in Table~\ref{tab.core}, and we note that  all these targets have a shared essentiality of at least 0.56. 
It is then tempting to suppose that the set of strictly retained genes is a reservoir of 
highly druggable genes, characterised both by highly shared orthology and essentiality, 
to be further exploited in the design of next generation antimicrobial drugs \citep{Chessher2012}. 
This result somehow specialises what \citet{Luo2015} found on the same set of bacterial species: 
``essential genes in the functional COG categories G, H, I, J, K and L tend to be more evolutionarily conserved 
than the corresponding nonessential genes in bacteria''. This kind of general statement 
deserves more investigation. First of all, in the next section we consider how 
essential and nonessential genes are partitioned into different COGs.

\begin{table*}[htbp]
\centering
\caption{{\bf List of strictly retained genes ($\mbox{ERI}=1$).} These 83 genes have orthologous copies in all the 45 bacteria of Table \ref{tab.dataset}, 
and are sorted by Z-score values of $K_a/K_s$. $f(\mbox{E})$ and $f(\mbox{NE})$ is the fraction of genomes in which each gene is annotated as essential and nonessential, respectively, 
and $f(\mbox{E})+f(\mbox{NE})=1$ only when the gene is fully annotated. Note the prevalence of COG J. In bold we report the genes which are antibiotic drug targets in \EC (see $https://www.drugbank.ca/$).}
\tiny
\begin{tabular}{|l|l|l|r|r|r|} 
\hline						
{\bf COG ID}	 &	{\bf Gene}  &{\bf Protein name} & {\bf f(E)} & {\bf  f(NE)}  & {\bf Z-score}  \\
\hline\hline
COG0238J	&	rpsR	&	30S ribosomal protein S18	&	0.60	&	0.28	&	-18.79	\\
COG0197J	&	rplP	&	50S ribosomal protein L16	&	0.80	&	0.16	&	-17.64	\\
COG0080J	&	rplK	&	50S ribosomal protein L11	&	0.76	&	0.12	&	-16.44	\\
COG0250K	&	-	&	transcription antitermination protein NusG	&	0.52	&	0.32	&	-16.13	\\
{\bf COG0185J}	&	rpsS	&	30S ribosomal protein S19	&	0.68	&	0.20	&	-15.80	\\
COG0100J	&	rpS11	&	30S ribosomal protein S11	&	0.76	&	0.12	&	-15.47	\\
{\bf COG0199J}	&	rpsN	&	30S ribosomal protein S14	&	0.64	&	0.24	&	-15.09	\\
COG0090J	&	rplB	&	50S ribosomal protein L2	&	0.80	&	0.12	&	-14.84	\\
{\bf COG0202K}	&	rpoA	&	DNA-directed RNA polymerase subunit alpha	&	0.80	&	0.12	&	-14.72	\\
COG0093J	&	rplN	&	50S ribosomal protein L14	&	0.72	&	0.16	&	-14.62	\\
{\bf COG0092J}	&	rpsC	&	30S ribosomal protein S3	&	0.88	&	0.04	&	-14.44	\\
COG0081J	&	rplA	&	50S ribosomal protein L1	&	0.64	&	0.24	&	-14.19	\\
{\bf COG0049J}	&	rpS7	&	30S ribosomal protein S7	&	0.88	&	0.08	&	-13.90	\\
COG0203J	&	rplQ	&	50S ribosomal protein L17	&	0.64	&	0.24	&	-13.74	\\
COG0211J	&	rpmA	&	50S ribosomal protein L27	&	0.72	&	0.12	&	-13.74	\\
COG0255J	&	rpmC	&	50S ribosomal protein L29	&	0.72	&	0.20	&	-13.66	\\
{\bf COG0103J}	&	rpsI	&	30S ribosomal protein S9	&	0.56	&	0.32	&	-13.20	\\
COG0231J	&	efp	&	elongation factor P		&	0.48	&	0.48	&	-12.75	\\
COG0691O	&	smpB	&	SsrA-binding protein		&	0.28	&	0.52	&	-12.21	\\
COG0201U	&	secY	&	preprotein translocase subunit SecY	&	0.84	&	0.08	&	-12.18	\\
COG0186J	&	rpsQ	&	30S ribosomal protein S17	&	0.72	&	0.16	&	-10.59	\\
COG0050J	&	tuf	&	elongation factor Tu		&	0.48	&	0.36	&	-10.43	\\
COG0184J	&	rpsO	&	30S ribosomal protein S15	&	0.52	&	0.36	&	-9.96	\\
COG0305L	&	dnaB	&	replicative DNA helicase	&	0.80	&	0.16	&	-9.07	\\
COG0336J	&	trmD	&	tRNA (guanine-N(1)-)-methyltransferase	&	0.60	&	0.28	&	-9.07	\\
COG0261J	&	rplU	&	50S ribosomal protein L21	&	0.56	&	0.28	&	-8.87	\\
COG0858J	&	-	&	ribosome-binding factor A	&	0.40	&	0.36	&	-8.70	\\
COG0781K	&	-	&	transcription termination/antitermination protein NusB	&	0.48	&	0.40	&	-8.46	\\
COG0097J	&	rplF	&	50S ribosomal protein L6	&	0.88	&	0.08	&	-7.18	\\
COG0222J	&	rplL	&	50S ribosomal protein L7/L12	&	0.68	&	0.20	&	-6.97	\\
{\bf COG0088J}	&	rplD	&	50S ribosomal protein L4	&	0.84	&	0.04	&	-6.33	\\
COG0089J	&	rplW	&	50S ribosomal protein L23	&	0.72	&	0.20	&	-5.60	\\
COG0091J	&	rplV	&	50S ribosomal protein L22	&	0.72	&	0.20	&	-5.58	\\
COG0087J	&	rplC	&	50S ribosomal protein L3	&	0.84	&	0.08	&	-5.26	\\
COG0541U	&	ffh	&	signal recognition particle protein	&	0.80	&	0.12	&	-4.82	\\
COG0592L	&	dnaN	&	DNA polymerase III subunit beta	&	0.72	&	0.16	&	-4.47	\\
COG0236IQ	&	-	&	acyl carrier protein		&	0.48	&	0.32	&	-3.67	\\
COG0480J	&	fus	&	elongation factor G		&	0.56	&	0.40	&	-2.87	\\
COG0233J	&	frr	&	ribosome recycling factor	&	0.88	&	0.12	&	-2.80	\\
COG0335J	&	rplS	&	50S ribosomal protein L19	&	0.56	&	0.32	&	-2.48	\\
COG0200J	&	rplO	&	50S ribosomal protein L15	&	0.72	&	0.16	&	-2.41	\\
COG0084L	&	-	&	TatD family deoxyribonuclease	&	0.16	&	0.80	&	-2.19	\\
COG0264J	&	tsf	&	elongation factor Ts		&	0.80	&	0.12	&	-2.15	\\
COG0528F	&	pyrH	&	uridylate kinase		&	0.84	&	0.16	&	-1.42	\\
COG0525J	&	valS	&	valyl-tRNA synthetase		&	0.84	&	0.08	&	-1.07	\\
COG0359J	&	rplI	&	50S ribosomal protein L9	&	0.24	&	0.68	&	-0.42	\\
COG1136V	&	-	&	ABC transporter ATP-binding protein	&	0.20	&	0.76	&	-0.20	\\
COG1136V	&	-	&	ABC transporter ATP-binding protein	&	0.20	&	0.76	&	-0.20	\\
COG1136V	&	-	&	ABC transporter ATP-binding protein	&	0.20	&	0.76	&	-0.20	\\
COG0172J	&	serS	&	seryl-tRNA synthetase		&	0.84	&	0.12	&	0.29	\\
COG0012J	&	-	&	GTP-dependent nucleic acid-binding protein EngD	&	0.08	&	0.84	&	0.59	\\
COG0571K	&	rnc	&	ribonuclease III		&	0.24	&	0.60	&	1.13	\\
COG0563F	&	adk	&	adenylate kinase		&	0.64	&	0.32	&	1.39	\\
COG0244J	&	rplJ	&	50S ribosomal protein L10	&	0.72	&	0.16	&	1.71	\\
{\bf COG0522J}	&	rpsD	&	30S ribosomal protein S4	&	0.80	&	0.12	&	2.03	\\
{\bf {\bf COG0188L}}	&	gyrA	&	DNA gyrase subunit A	&	0.64	&	0.24	&	2.27	\\
{\bf {\bf COG0188L}}	&	gyrA	&	DNA gyrase subunit A	&	0.64	&	0.24	&	2.27	\\
{\bf COG0086K}	&	rpoC	&	DNA-directed RNA polymerase subunit beta	&	0.76	&	0.20	&	2.36	\\
COG0180J	&	trpS	&	tryptophanyl-tRNA synthetase	&	0.72	&	0.20	&	2.61	\\
COG0629L	&	ssb	&	single-strand binding protein family	&	0.48	&	0.44	&	2.65	\\
COG0195K	&	nusA	&	transcription elongation factor NusA	&	0.80	&	0.12	&	2.94	\\
COG0013J	&	alaS	&	alanyl-tRNA synthetase		&	0.68	&	0.28	&	3.70	\\
COG0228J	&	rpsP	&	30S ribosomal protein S16	&	0.64	&	0.28	&	3.88	\\
COG0102J	&	rplM	&	50S ribosomal protein L13	&	0.80	&	0.04	&	4.57	\\
COG0568K	&	rpoD	&	RNA polymerase sigma factor	&	0.64	&	0.36	&	4.58	\\
COG0072J	&	pheT	&	phenylalanyl-tRNA synthetase subunit beta	&	0.76	&	0.20	&	4.75	\\
COG0544O	&	tig	&	trigger factor			&	0.00	&	0.88	&	5.11	\\
COG0209F	&	nrdE	&	ribonucleotide-diphosphate reductase subunit alpha	&	0.64	&	0.32	&	5.34	\\
COG0484O	&	-	&	DnaJ domain-containing protein	&	0.40	&	0.56	&	5.46	\\
COG0484O	&	-	&	DnaJ domain-containing protein	&	0.40	&	0.56	&	5.46	\\
COG0484O	&	-	&	DnaJ domain-containing protein	&	0.40	&	0.56	&	5.46	\\
COG0443O	&	dnaK	&	molecular chaperone DnaK	&	0.60	&	0.36	&	5.82	\\
COG0442J	&	proS	&	prolyl-tRNA synthetase		&	0.92	&	0.08	&	5.91	\\
COG0018J	&	argS	&	arginyl-tRNA synthetase		&	0.72	&	0.20	&	9.22	\\
COG0215J	&	cysS	&	cysteinyl-tRNA synthetase	&	0.88	&	0.08	&	9.42	\\
COG0536R	&	obgE	&	GTPase ObgE			&	0.68	&	0.24	&	11.84	\\
COG0552U	&	ftsY	&	signal recognition particle-docking protein FtsY	&	0.80	&	0.12	&	12.44	\\
COG0576O	&	grpE	&	co-chaperone GrpE		&	0.68	&	0.24	&	13.17	\\
COG0143J	&	metS	&	methionyl-tRNA synthetase	&	0.64	&	0.32	&	14.64	\\
COG1214O	&	-	&	glycoprotease			&	0.60	&	0.32	&	15.12	\\
COG0587L	&	polC-2	&	DNA polymerase III subunit alpha	&	0.72	&	0.24	&	20.79	\\
COG0223J	&	fmt	&	methionyl-tRNA formyltransferase	&	0.68	&	0.24	&	25.68	\\
COG0313R	&	-	&	tetrapyrrole (corrin/porphyrin) methylase protein	&	0.08	&	0.76	&	26.42	\\
\hline
\end{tabular}
\label{tab.core}
\end{table*}

\begin{table*}[htbp]
\centering
\caption{{\bf Functional specialization of essential and nonessential genes according to COG clusters.} 
Figures indicate the percentages of essential and nonessential genes within a given COG (sums of these figures for table subsections are reported in boldface). COGs are sorted by percent essentiality.}
\footnotesize
\begin{tabular}{clrr} 		
\hline\hline			
{\bf COG ID}& {\bf Functional classification}	 &	{\bf \% E } & {\bf \% NE}    \\
\hline
& INFORMATION STORAGE AND PROCESSING & &\\
J & Translation, ribosomal structure and biogenesis &0.25\ & 0.05\\
K &	Transcription &0.06 & 0.10\\ 
L & Replication, recombination and repair	&0.08 & 0.07\\ 
& & {\bf 0.39}& {\bf 0.22}\\
\hline
&  CELLULAR PROCESSES AND SIGNALING & &\\
D &	Cell cycle control, cell division, chromosome partitioning &0.03 & 0.01\\ 
T & Signal transduction mechanisms	&0.02 & 0.07\\
M &	Cell wall/membrane/envelope biogenesis &0.10 & 0.08\\
N & Cell motility&0.01 & 0.03\\
O & Posttranslational modification, protein turnover, chaperones	&0.04 & 0.05\\ 
& & {\bf 0.20}& {\bf 0.24}\\
\hline
& METABOLISM & &\\
C & Energy production and conversion &0.07 & 0.08\\ 
G & Carbohydrate transport and metabolism &0.06 & 0.10\\
E &	Amino acid transport and metabolism &0.06 & 0.12\\ 
F & Nucleotide transport and metabolism	&0.05 & 0.03\\
H & Coenzyme transport and metabolism	&0.08 & 0.06\\
I & Lipid transport and metabolism &0.07& 0.05\\
P &	Inorganic ion transport and metabolism &0.03 & 0.08\\
& & {\bf 0.41}& {\bf 0.52}\\
\hline\end{tabular}
\label{tab.func}
\end{table*}

\subsection{Functional specialization of essential and nonessential genes}
\begin{figure}
 \includegraphics[width=8.6cm]{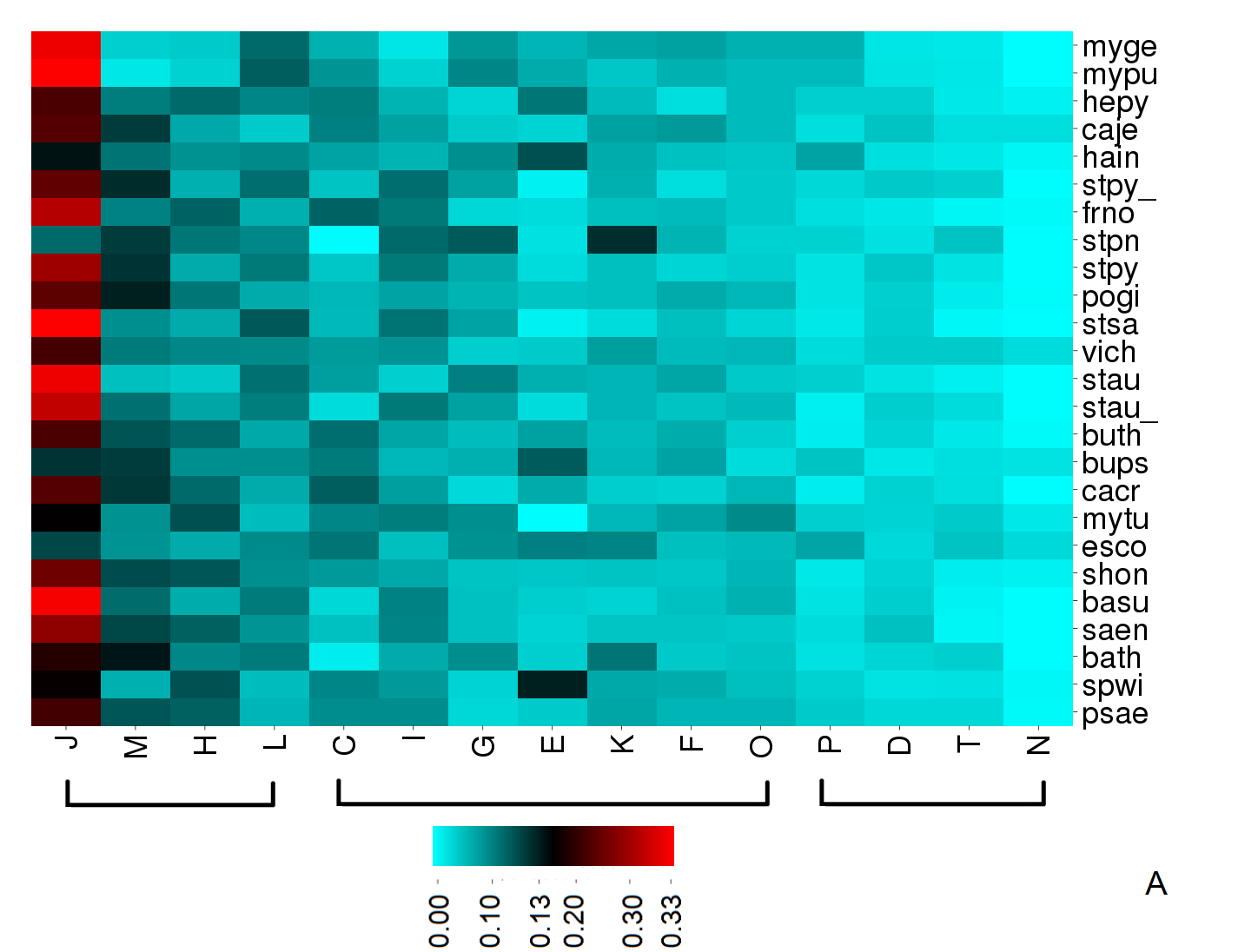}
 \includegraphics[width=8.6cm]{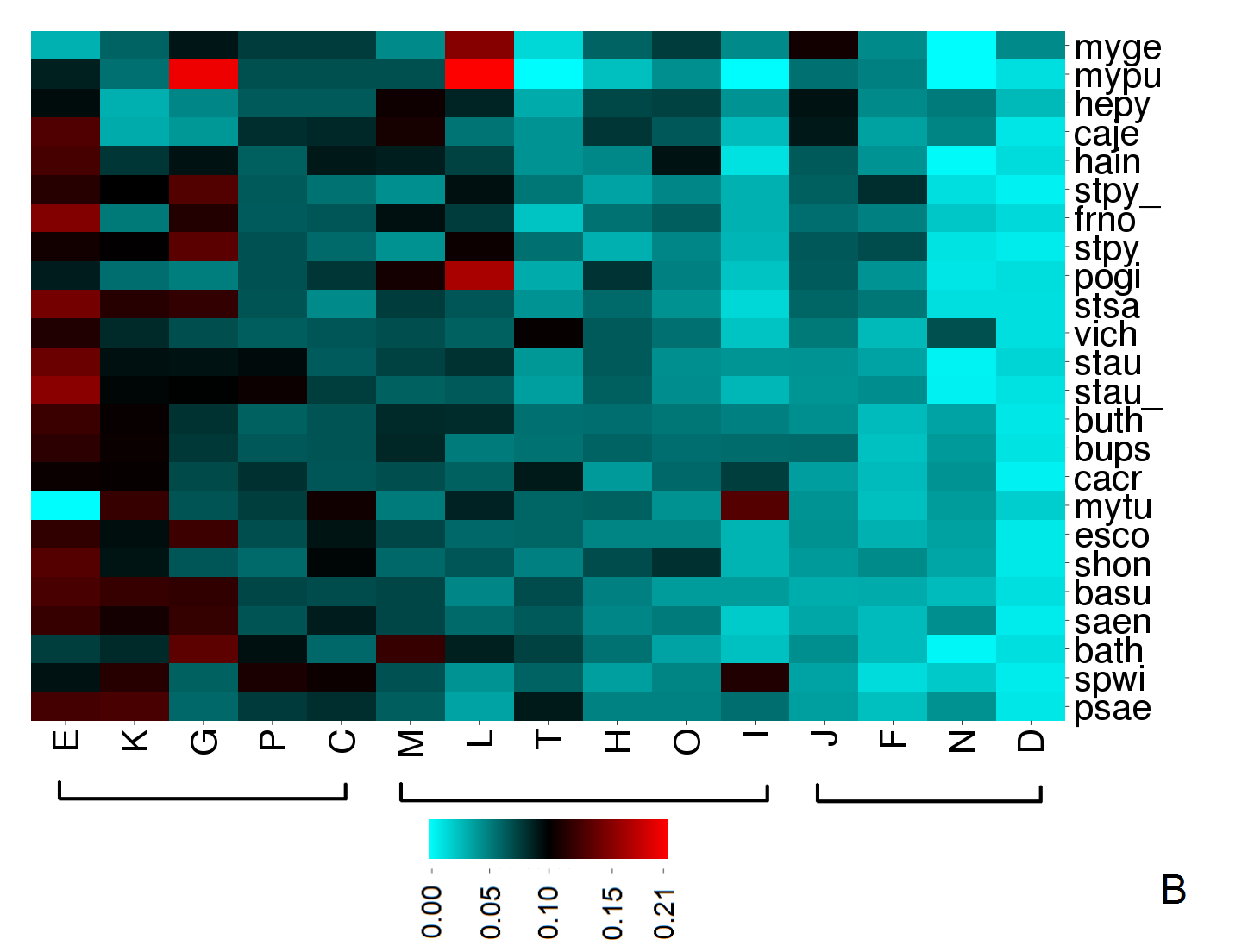}
\caption{For each genome we estimated the conditional probabilities $P(\mbox{COG}|\mbox{E})$ (panel A) 
and $P(\mbox{COG}|\mbox{NE})$ (panel B) for an essential and a nonessential gene to belong to a given COG. 
Genomes are ranked from top to bottom by the size of their genomes. 
COGs are ranked, separately in both panels and from left to right, 
according to their overall incidence. In panel A, 51\% of the essential genes 
belong, in different proportions, to COGs J, M, H and L; 40\% to C, I, G, E, K, 
F and O and the remaining 10\% to  P, D, T and N. In panel B, 49\% of 
the nonessential genes belong to E, K, G, P, and C; 38\% to  M, L,T, H, O and 
I; the remaining 13\% to J, F, N, and D.}
\label{fig3}      
\end{figure}

The heat maps of Figure~\ref{fig3} represent conditional probabilities $P(\mbox{COG}|\mbox{E})$ and $P(\mbox{COG}|\mbox{NE})$ that 
essential and nonessential genes belong to the different COGs, for the various 
bacterial species we consider. Essential and nonessential genes have different 
functional spectra. In both panels, a banded vertical structure emerges which roughly 
separates COGs into three groups. In particular, 51\% of essential genes fall 
into J, M, H and L , whereas 49\% of nonessential genes belong to E, K, G, P and C. 
Table~\ref{tab.func} synthetically shows that essential genes dominate functions 
related to {\it information storage and processing}, whereas nonessential genes 
prevail among the set of functions related to {\it metabolism}. Functions 
related to {\it cellular processes and signalling} appear to be equally shared between essential and nonessential genes. 
In the next section, using the criteria of the $K_a/K_s$ ratio,  we challenge the sensible statement that essential genes 
are subject to a stricter purifying selection than nonessential genes. If that were true, then each COG 
would exhibit a signature of either purifying or positive selection, on the basis of the fraction of essential genes that belong to it.

\subsection{Selective pressure, conservation and essentiality}

\begin{figure}
\includegraphics[width=8.6cm]{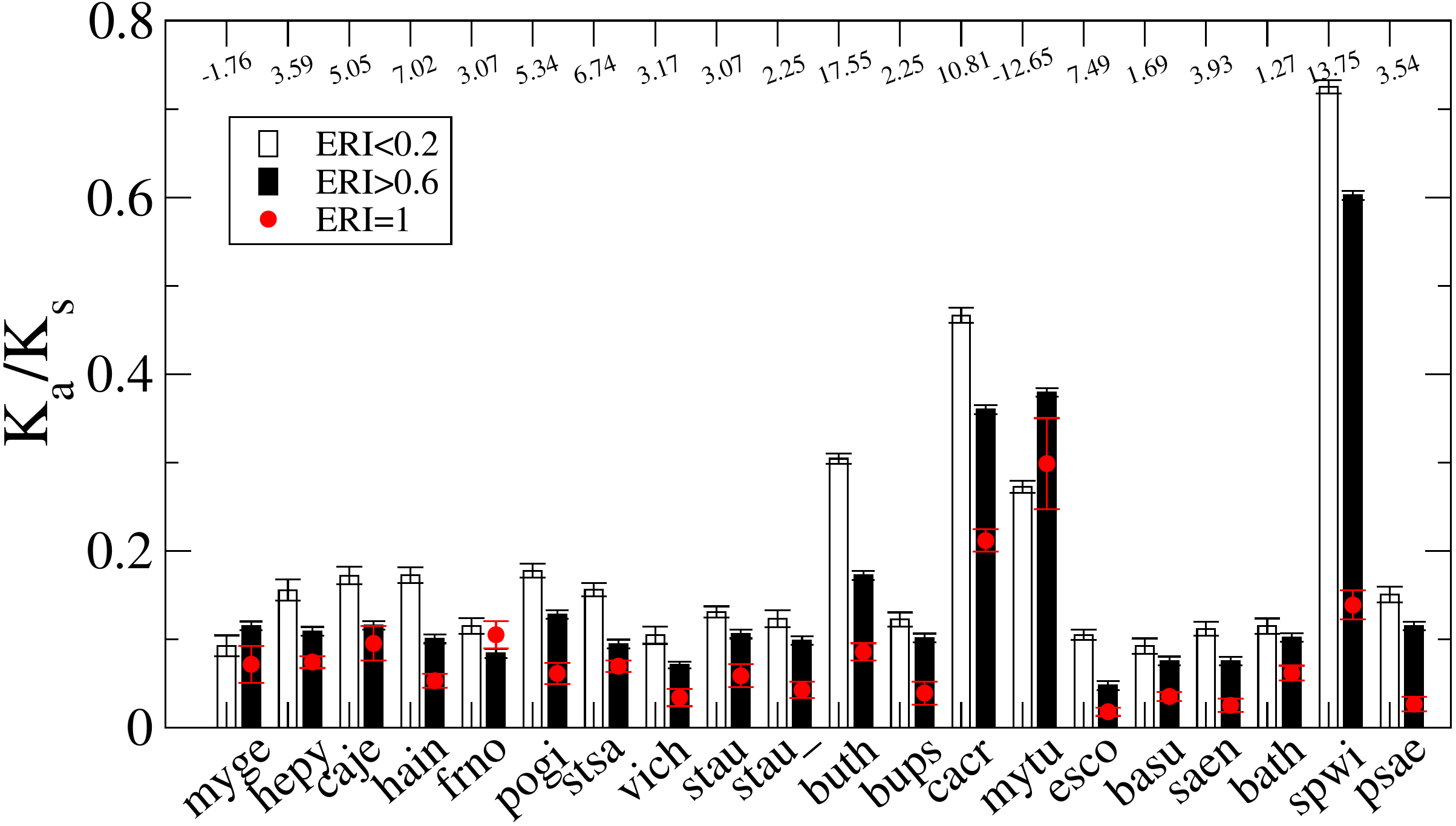}
\caption{Average values and standard deviations of $K_a/K_s$ for specific and conserved genes in each genome. Specific genes have $\mbox{ERI}<0.2$ and conserved genes have $\mbox{ERI}>0.6$. 
According to a two-sample z statistic, whose values are shown at the top of the panel, with the exception of $myge$ and $mytu$, the average value of $K_a/K_s$ is significantly higher for specific genes. 
Red points denote averages and errors of $K_a/K_s$ for the most conserved genes (those with $\mbox{ERI}=1$).}
\label{fig4}      
\end{figure}

In this section we firstly consider how evolutionary pressure, as represented by 
the ratio $K_a/K_s$, correlates with the degree of retention (conservation) of 
bacterial genes. Note that each bacterial genome has its own level of evolutionary 
pressure (see Figures~\ref{fig4} and~\ref{fig10}). We thus compare, within each genome, the 
evolutionary pressure that is exerted over more or less conserved genes. 
Using the thresholds of ERI used in \citet{Dilucca2015}, Figure~\ref{fig4} shows that 
more conserved genes (with $\mbox{ERI}>0.6$) significantly display 
lower values of $K_a/K_s$ than less conserved genes (with $\mbox{ERI}<0.2$). Interestingly, 
genes belonging to the core of 83 strictly conserved genes of $\mbox{ERI}=1$, mentioned above, 
have levels of $K_a/K_s$ that are systematically below the average value of the more conserved genes. 
This last observation stresses once more that the most conserved genes, those involved 
in more basic and universal functions, tend to be subject to a relatively purifying, conservative selection. 
Since highly conserved genes are also prone to be essential, as shown in Figure~\ref{fig1}, 
our observation confirms the previous conclusion by \citet{Luo2015} that 
``essential genes are more evolutionarily conserved (they are characterised 
by a significantly lower $K_a/K_s$) than nonessential ones in most of the bacteria''.

\begin{figure}
\includegraphics[width=8.6cm]{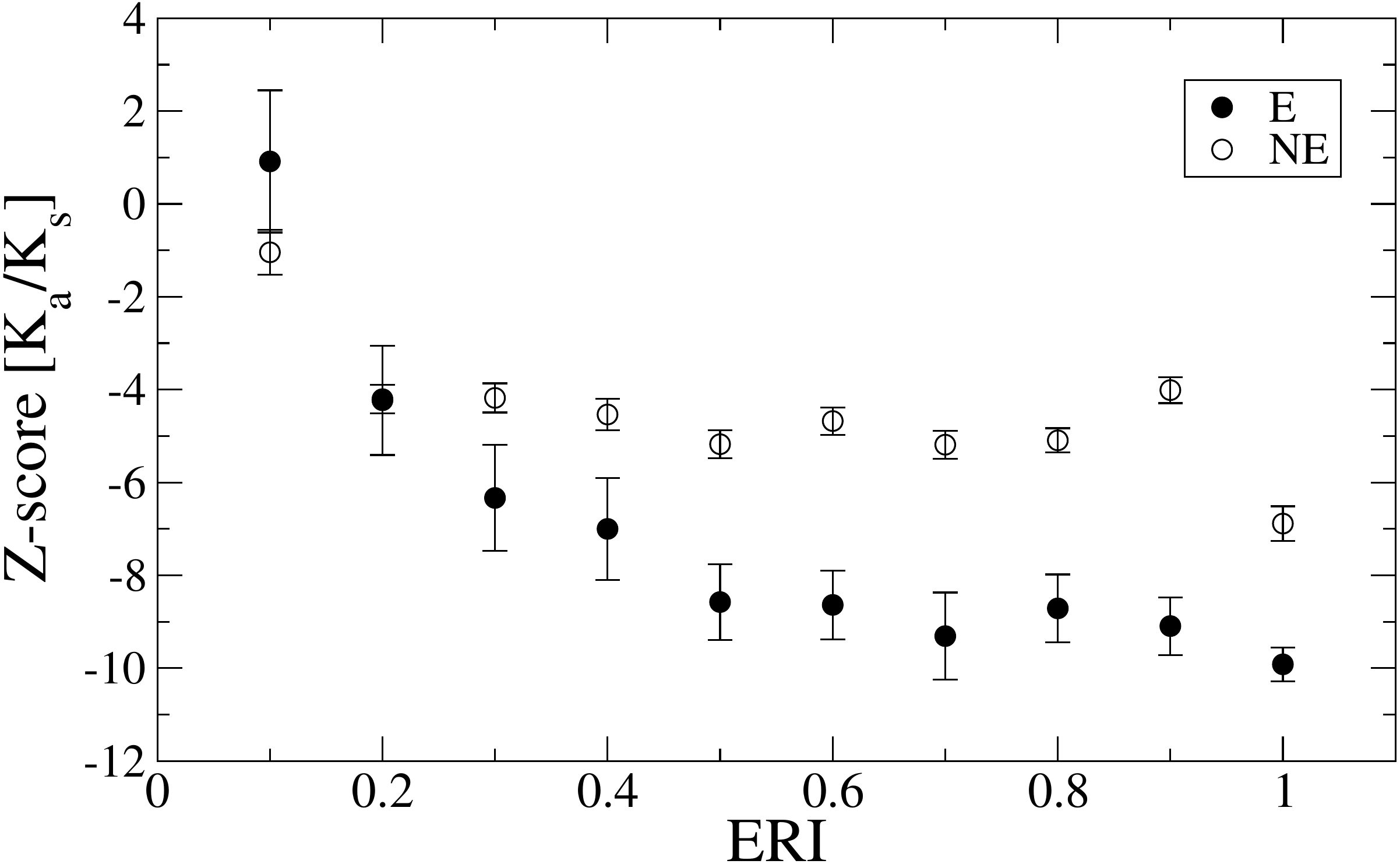}
\caption{Z-scores of $K_a/K_s$, relative to the average value in the species, 
for essential and nonessential genes within binned ERI values, for the DEG-annotated genomes of Table~\ref{tab.dataset}. 
Error bars are root mean square deviations for the genes falling in each bin. 
Z-scores decrease with the degree of conservation: the more a gene is retained among different species, 
the more is subject to a purifying selection. The two trends are well separated 
(with the exception of less conserved genes): average Z-scores of essential genes are systematically lower than those of nonessential genes, 
confirming that essential genes are subject to a more purifying, conservative evolutionary pressure.}
\label{fig5}      
\end{figure}

Looking for a general relationship between the evolutive pressure exerted on a 
gene and its degree of conservation as measured by the ERI, in Figure~\ref{fig5} 
we show that when the degree of retention increases, the Z-score of $K_a/K_s$ 
systematically decreases, becoming more and more negative. This 
observation stresses again that those genes which are common to several species 
are subject to a purifying, more constrained evolution. Note that, 
comparatively, essential genes systematically have a Z-score which is more 
negative than for nonessential genes, indicating that they are, for each 
degree of retention, subject to a more purifying evolution.

\begin{figure}
  \includegraphics[width=8.6cm]{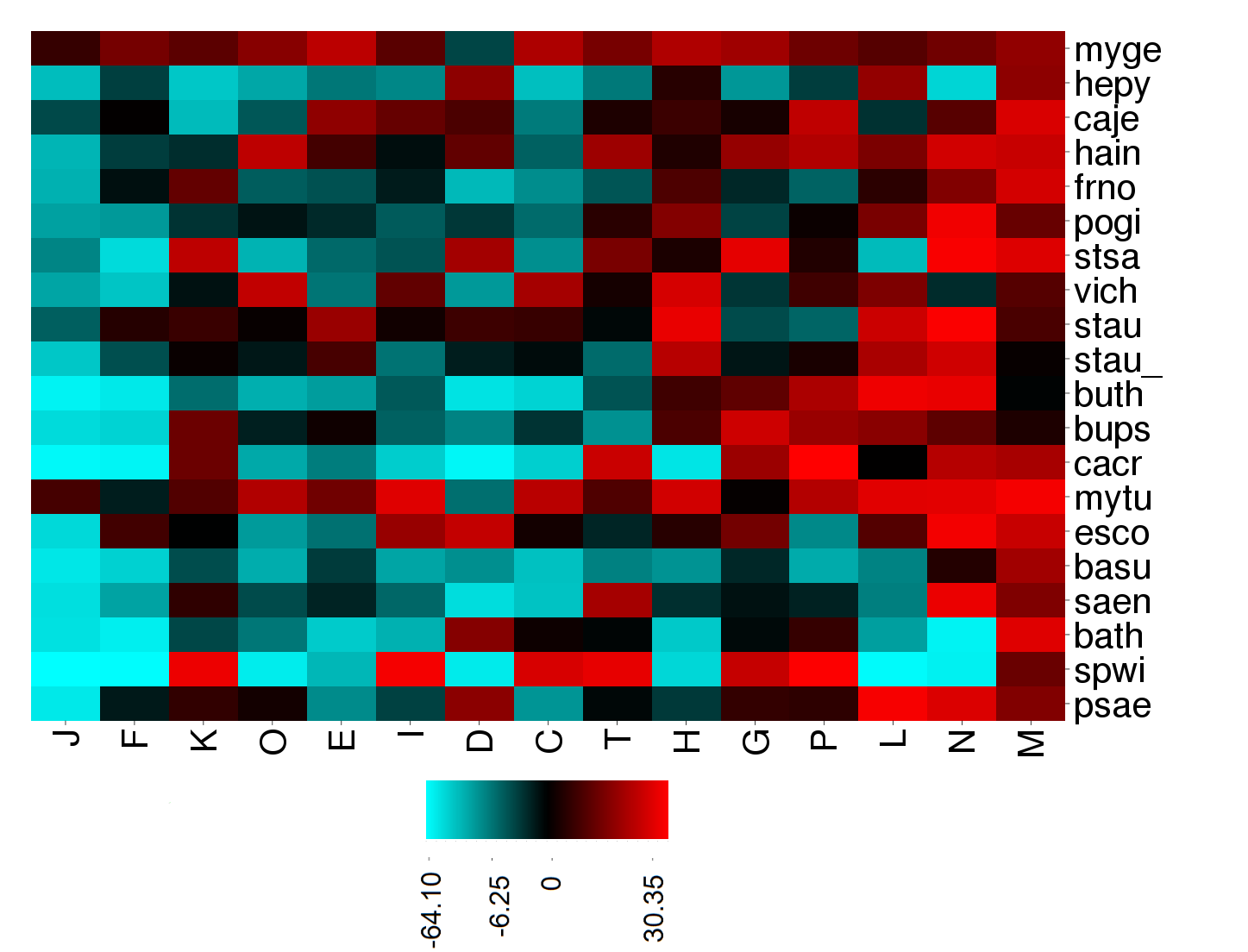}
\caption{Z-score of the $K_a/K_s$ ratios for each COG, with respect to the average value in each bacteria. 
In the color scale, red means significant positive Z-score (selective pressure), 
whereas, green indicates significant negative Z-score (purifying pressure). 
Bacteria are ordered according to their genome size (from top to bottom), while 
COGs are ordered from left to right according to the ranking of Table~\ref{tab.zeta}}
\label{fig6}      
\end{figure}

Getting to the functional annotation provided by COGs, \citet{Luo2015} also show 
that essential genes in each of the COGs G, H, I, J, K and L tend to be significantly more 
evolutionarily conserved than nonessential genes belonging to the same COGs. 
It would be then natural to conclude that the nature of the evolutionary 
pressure that is exerted on the genes belonging to a COG depends on its content 
of essential genes. From Table~\ref{tab.func} it is possible to rank bacterial COGs 
and their functions by their content of essential genes. In particular, note 
that COGs J, M, H and L contain more than 51\% of the annotated essential genes, 
and a recent experiment has re-confirmed in $basu$ that precisely 
these COGs are enriched in essential genes (see Figure 2 in \citet{Koo2017}).
One would then conclude that the genes in these COGs should be under a more 
conservative evolutionary pressure than those belonging to the rest of the COGs. 
To elucidate this point, we evaluated the Z-scores of $K_a/K_s$ over the genes 
of each COG with respect to the average value of this ratio over the genomes 
they come from (results in Table~\ref{tab.zeta} and in Figure~\ref{fig6}). 
According to this analysis, one would conclude that the rank order 
of the evolutionary pressure on the COGs would be J, F, K, O, E, I, D, C, T, H, G, P, L, N, M, 
going from relatively purifying to more diversifying selection. 
In the first four ranks we find J, F, K and O, 
at variance with the expectation based on the content of essential genes.

From one point of view one could think that the observed discrepancy between the 
ranking based on the Z-score and the one based on the essentiality content should 
depend on the limited coverage of the available dataset. At present not all 
genes in the genomes we have investigated are annotated for essentiality and, 
even-worse, not all the genes have been attributed to a COG class (see Table \ref{tab.dataset} 
to check for the coverage of the essentiality and COG annotation). From another point of view, 
our Z-score statistics in Figure \ref{fig6} is based on a set of 39.804 genes (annotated for $K_a/K_s$ and COG) over a total of 127.012. 
We believe our Z-score statistics is sufficiently representative of the overall 
evolutionary pressure exerted over the COG classes. On the basis of the data in 
Figure~\ref{fig6}, we propose a tentative distinction between a set of 
relatively more evolutionarily conserved COGs (J, F, K, and O)  and, on the 
other side, a set of more adaptive ones (P, L, N, and M). This distinction 
should be further tested, along with the progressive annotation of bacterial genomes. 
Notably COG J, the set of genes which is more exhaustively annotated, has the highest 
percentage of essential genes and with little uncertainty is, in all genomes, 
under a conservative evolutionary pressure. 

\begin{figure}
  \includegraphics[width=8.6cm]{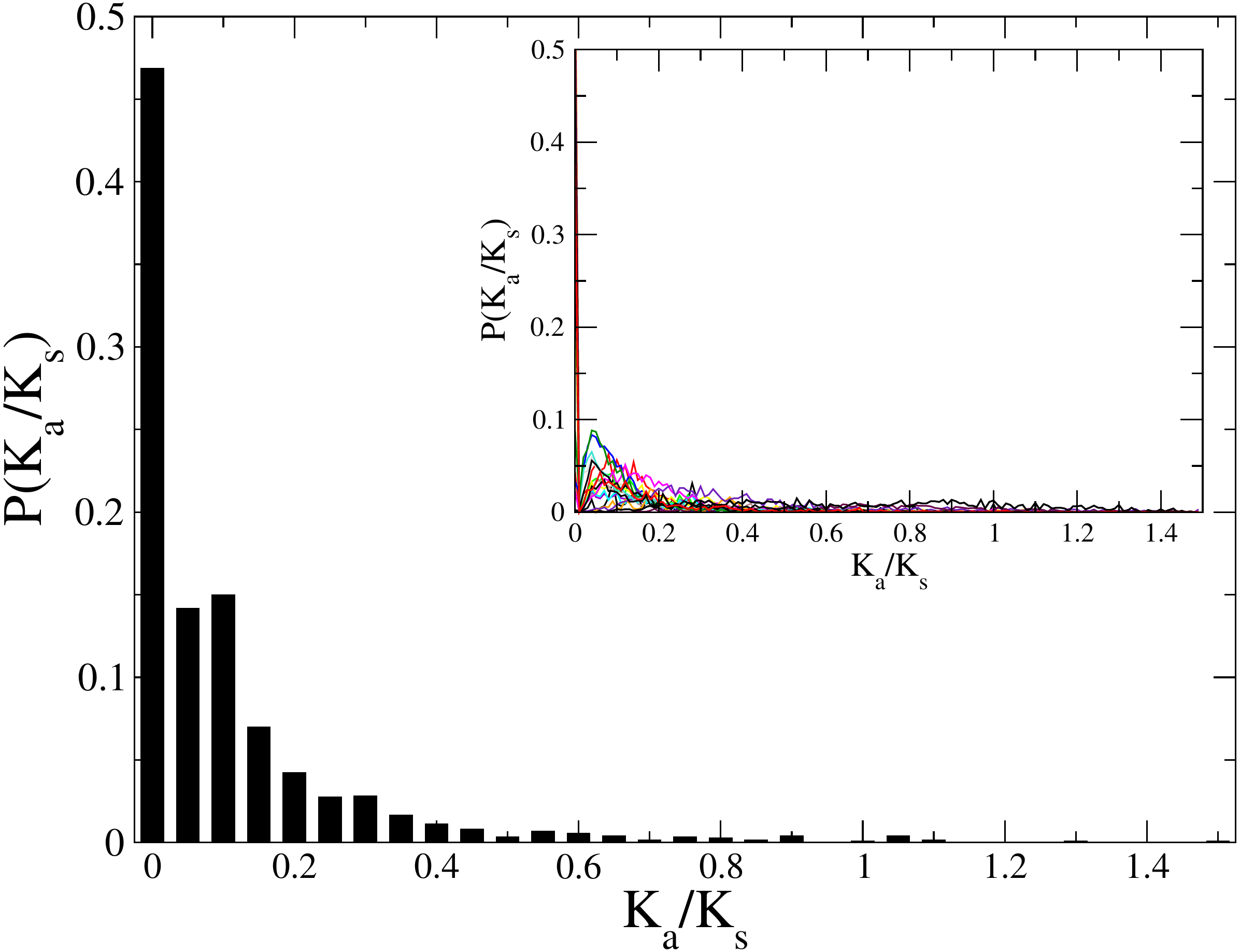}
\caption{Frequency distribution of $K_a/K_s$ for the 83 genes with $\mbox{ERI}=1$, those having orthologous in all the genomes of the DEG-annotated species of Table~\ref{tab.dataset}. 
These genes tend to display values of $K_a/K_s$ concentrated between 0 and 0.2. 
Indeed, the average value is 0.085, the standard deviation is 0.004 and the median is 0.070, 
and only in very few cases values bigger than 1 are observed. 
Inset: same distributions for all genes within individual bacterial genomes}
\label{fig7}      
\end{figure}

To conclude this section we focus on the set of 83 genes 
which are highly retained in all species (those with $\mbox{ERI}=1$) and which display a restricted 
set of functional specialisations. In Figure~\ref{fig7} we show the histogram 
of the $K_a/K_s$ values for this core of genes in all DEG-annotated species of Table~\ref{tab.dataset}. 
The plot indicates that even the genes with orthologous copies in all the genomes 
here considered rarely have values of $K_a/K_s$ bigger than 1. This shows that 
they generally are under an overall purifying selection. It is worth to note 
that this distribution of $K_a/K_s$ is consistent with the distributions of the same parameter 
over the different COGs. This last observation points out that a sufficiently large set of bacterial genes display a similar 
distribution of $K_a/K_s$, which in turn indicates that in general these genes 
are subject to an overall purifying selection ($K_a/K_s<1$). 
Nevertheless, through the relative comparison of the individual Z-scores of 
genes in different genomes and in different COGs, we can sensibly assess that 
the different COGs are under diverse evolutive pressure and constraints.

\begin{table}[h!]
\centering
\caption{{\bf Ranking of COGs according to Z-scores of $\mathbf{K_a/K_s}$.} We count $\pm1$ for each time the Z-score is smaller than -1 or bigger than +1, respectively.}
\footnotesize
\begin{tabular}{|c|c|c|c|c|c|c|c|c|c|c|c|c|c|c|c|} 
\hline
{\bf COG ID}	 &	J	&	F	&	K	&	O	&	E	&	I	&	D	&	C	&	T	&	H	&	G	&	P	&	L	&	N	&	M	\\
\hline
{\bf Z-score }   &	-20	&	-19	&	-14	&	-13	&	-13	&	-13	&	-12	&	-12	&	-9	&	-9	&	-6	&	-5	&	-5	&	4	&	5	\\
\hline
\end{tabular}
\label{tab.zeta}
\end{table}

\subsection{Codon bias patterns in bacterial genes} 

\begin{figure*}
  \includegraphics[width=12.9cm]{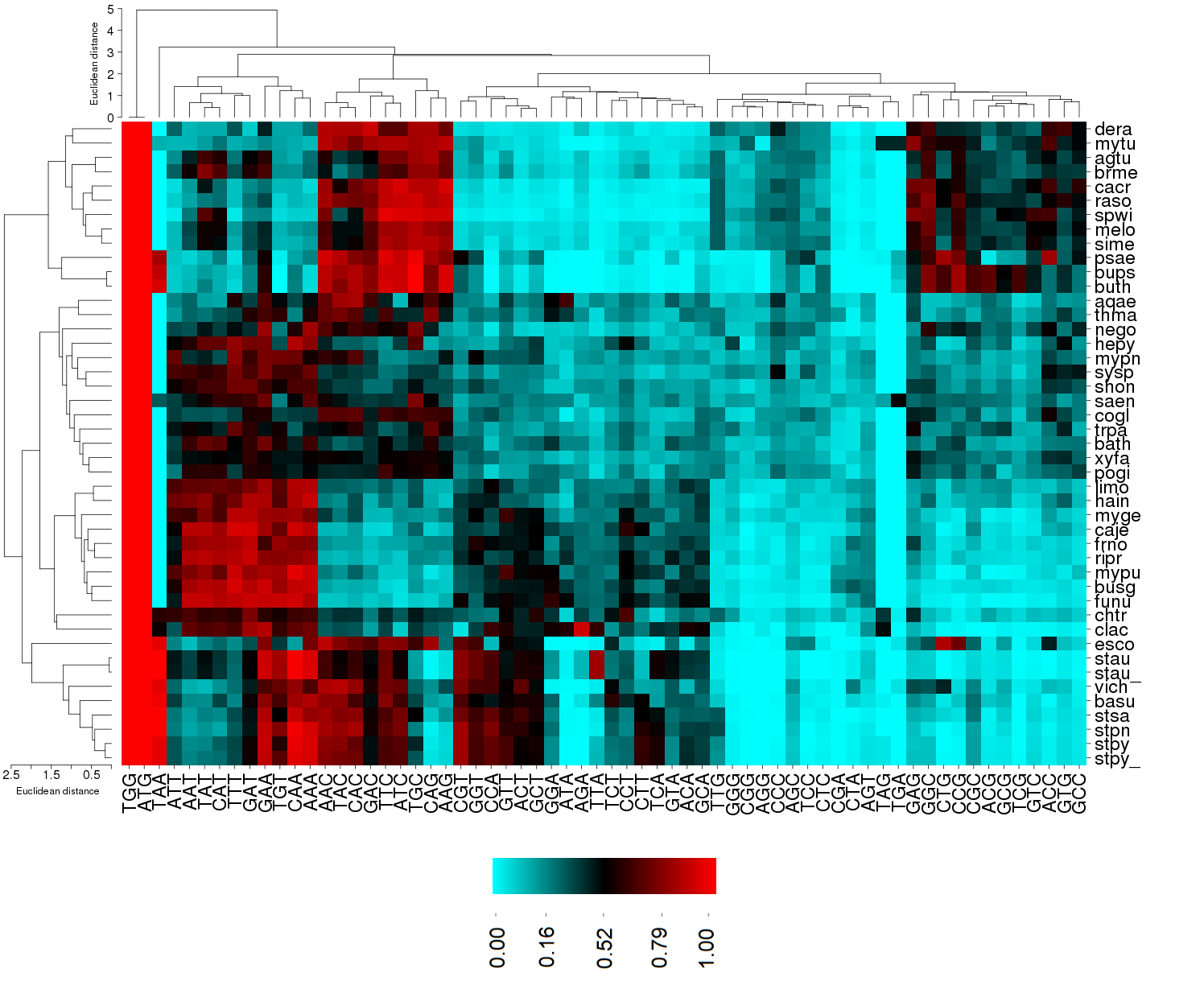}
  \begin{minipage}{0.38\textwidth}
   \includegraphics[width=6.45cm]{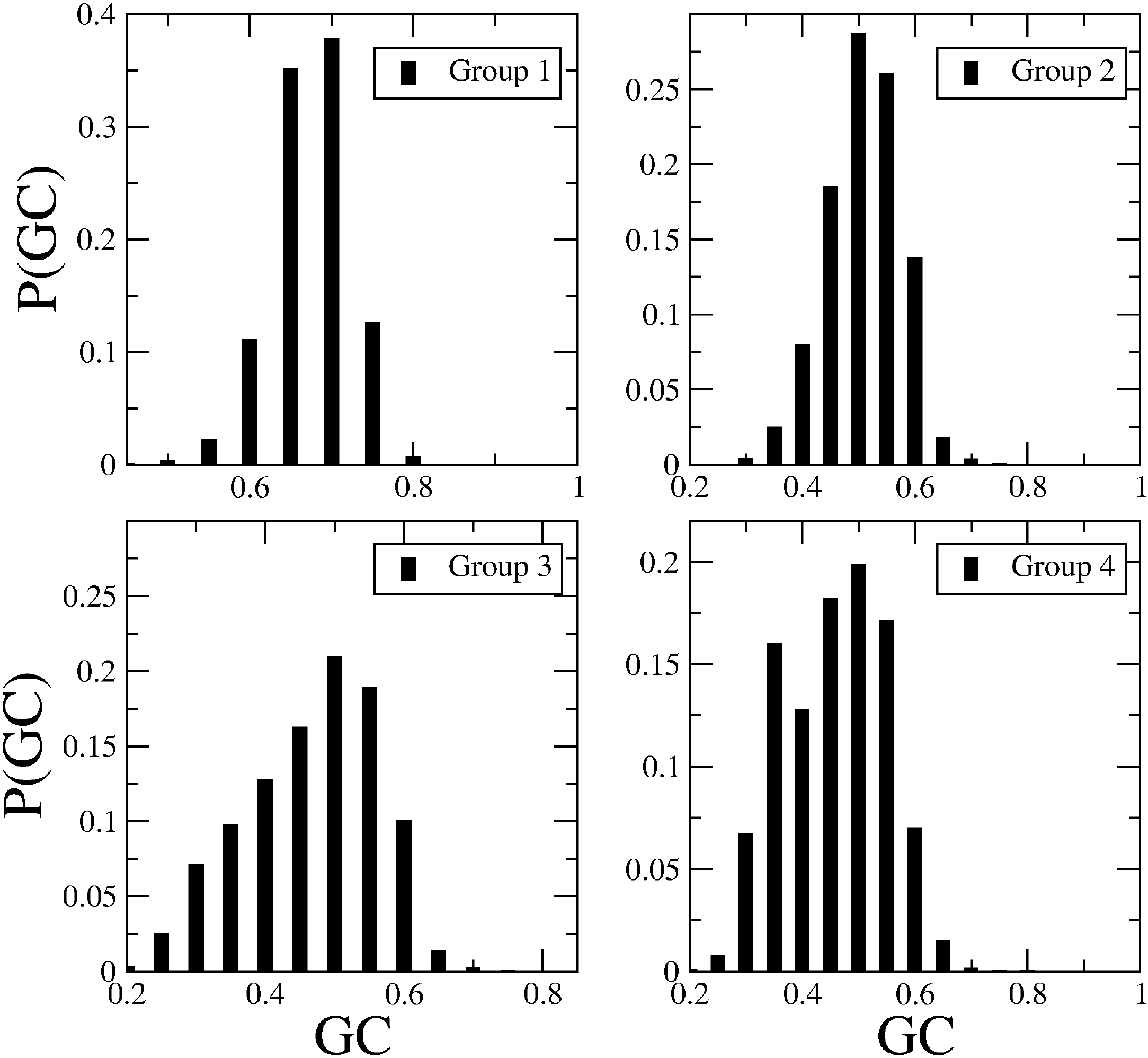}
  \end{minipage}
  \begin{minipage}{0.3\textwidth}
\tiny
{\bf Group 1} \\
Deinococcus radiodurans R1	\\
Mycobacterium tuberculosis H37Rv		\\
Agrobacterium tumefaciens (fabrum)	\\
Brucella melitensis bv. 1 str. 16M	\\	
Caulobacter crescentus		\\
Ralstonia solanacearum GMI1000	\\
Sphingomonas wittichii RW1		\\
Mesorhizobium loti MAFF303099	\\
Sinorhizobium meliloti 1021		\\
Pseudomonas aeruginosa UCBPP-PA14	\\
Burkholderia pseudomallei K96243	\\
Burkholderia thailandensis E264\\
\medskip
{\bf Group 3} \\
Listeria monocytogenes EGD-e	\\
Haemophilus influenzae Rd KW20	 \\
Mycoplasma genitalium G37		\\
Campylobacter jejuni 		\\
Francisella novicida U112	\\
Rickettsia prowazekii str. Madrid E \\
Mycoplasma pulmonis UAB CTIP	\\
Buchnera aphidicola Sg uid57913	\\
Fusobacterium nucleatum ATCC 25586
\end{minipage}
\begin{minipage}{0.3\textwidth}
\tiny
{\bf Group 2} \\
Aquifex aeolicus VF5	\\
Thermotoga maritima MSB8		\\
Neisseria gonorrhoeae FA 1090 uid57611 \\
Helicobacter pylori 26695 	\\
Synechocystis sp. PCC 6803		\\
Shewanella oneidensis MR-1		\\
Mycoplasma pneumoniae M129	\\
Salmonella enterica serovar Typhi		\\
Corynebacterium glutamicum ATCC 13032		\\
Treponema pallidum Nichols	\\
Bacteroides thetaiotaomicron VPI-5482	\\
Xylella fastidiosa 9a5c			\\
Porphyromonas gingivalis ATCC 33277 \\
\medskip
{\bf Group 4} \\
Bacillus subtilis 168	\\
Chlamydia trachomatis D/UW-3/CX \\
Clostridium acetobutylicum ATCC 824	\\
Escherichia coli K-12 MG1655		\\
Staphylococcus aureus N315	\\
Staphylococcus aureus ncTC 8325 \\
Streptococcus pneumoniae TIGR4	\\
Streptococcus pyogenes MGAS5448	\\
Streptococcus pyogenes NZ131 \\
Streptococcus sanguinis	\\
Vibrio cholerae N16961
\end{minipage}
\caption{RSCU values of individual codons in different species. Both genomes and groups of codons are clustered by similarity of codon usage. 
Note that bacterial strains of $mypu$, $shon$, $stpy$ and $stpy\_$ are missing in the dataset of \citet{Luo2015} as well as in the figure. 
Lower panel: distribution of GC content for genomes in each of the four groups of bacteria listed on the right of the panel. 
Mann–Whitney U Tests on the median return 0.05 (group 1 -- group 2), 0.003 (group 1 -- group 3), 0.001 (group 1 -- group 4), 0.73 (group 2 -- group 3), 0.824 (group 2 -- group 4), 0.910 (group 3 -- group 4).}
\label{fig8}      
\end{figure*}

Previous observations (see \citet{Plotkin2011} and data therein) 
point to the fact that each bacterial species has a specific pattern and level 
of codon bias, which is strongly shared by all its genes; 
codon bias in specialized categories of genes appears to be just 
a modulation of the distinctive codon bias of the species. 
To check this statement, we compute $RSCU$ values of each codon 
for our set of bacterial genomes, and plot results in Figure~\ref{fig8}---where 
both codon bias patterns and genomes are clustered according to 
similarity in the codon usage. The emerging striped structure 
indicates that these bacteria cluster into at least four groups, 
characterised by different patterns of codon usage (as measured by $RSCU$). 
Note that this grouping is not driven solely by the GC content of the genomes, 
as only the first group turns out to be well separated by the others in terms of the group-specific GC content distribution. 
Overall, this is just a preliminary exploration suggesting that there should be a strong 
correlation between codon bias patterns of each species and his evolutionary 
history. Further work is needed, in our opinion, to search for hidden 
ecological determinants behind this rough classification based on basic codon bias.

Following this line of reasoning, we check whether the essentiality of a gene 
has a signature in its codon usage bias. For each of our DEG-annotated genome, 
we thus compute $RSCU$ values separately for essential and nonessential genes. 
With the exception of $buth$ and $stsa$, the two $RSCU$ vectors are very similar 
for each genome. This indicates that the change in codon bias induced 
by essentiality, if any, is weak with respect to the prevailing codon bias 
signature of the species.

\begin{figure}
  \includegraphics[width=8.6cm]{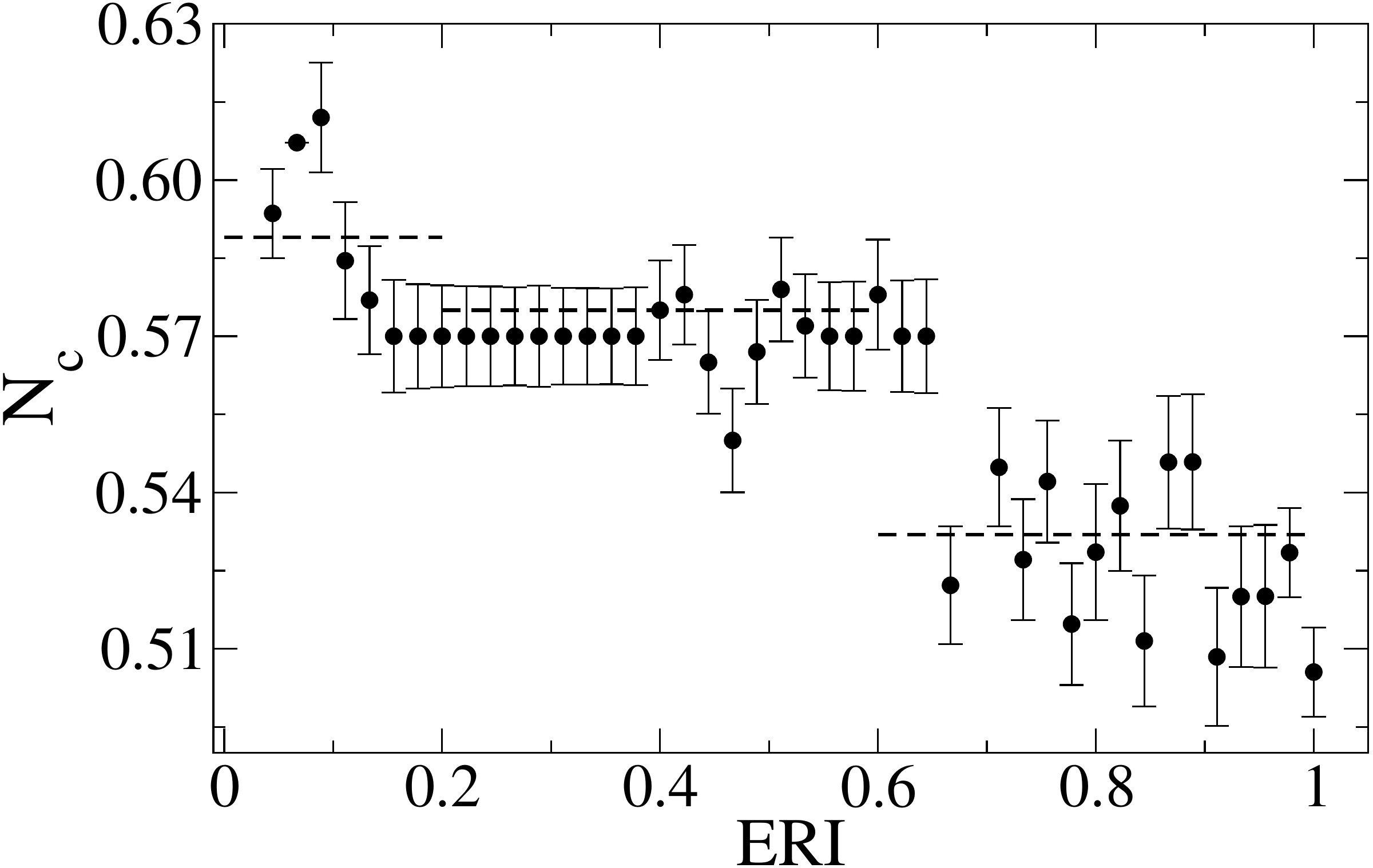}
\caption{Codon bias measured by $N_c$ for bacterial genes having similar values of ERI.
Note that $N_c$ values have significantly different averages among bacterial species \citep{Ran2012} 
and thus, for the sake of comparison, they have been normalised within each species between 0 and 1. 
The dashed lines represent average codon bias levels of genes in the groups of $\mbox{ERI}<0.2$ 
(specific genes), of $0.2<\mbox{ERI}<0.6$, and of $\mbox{ERI}>0.6$ (conserved genes).}
\label{fig9}      
\end{figure}

\subsection{Conservation and codon bias of bacterial genes} 

In order to investigate whether the codon bias of a bacterial gene is correlated 
with its degree of conservation, we plot in Figure~\ref{fig9} values of $N_c$ 
(normalised within the species) for genes with given values of ERI. 
As we did in \citet{Dilucca2015} for \EC, for each bacterial genome it is possible 
to separate groups of genes with different patterns of codon bias on 
the basis of their degree of conservation: those with $\mbox{ERI}<0.2$, those with 
$0.2<\mbox{ERI}<0.6$ and those with $\mbox{ERI}>0.6$. 
The evolutionary codon adaptation indeed tends to be higher for genes 
that are more conserved (genes with $\mbox{ERI}>0.6$ have lower values of $N_c$). 
Recall also from Figure~\ref{fig1} that groups of genes with $\mbox{ERI}<0.6$ 
have a probability of being essential that is less than 0.2. 
From these observation, we can conclude that the more a gene is conserved, 
the more it displays a selected choice of synonymous codons.

\subsection{Codon bias and evolutionary pressure}

\begin{figure}
  \includegraphics[width=8.6cm]{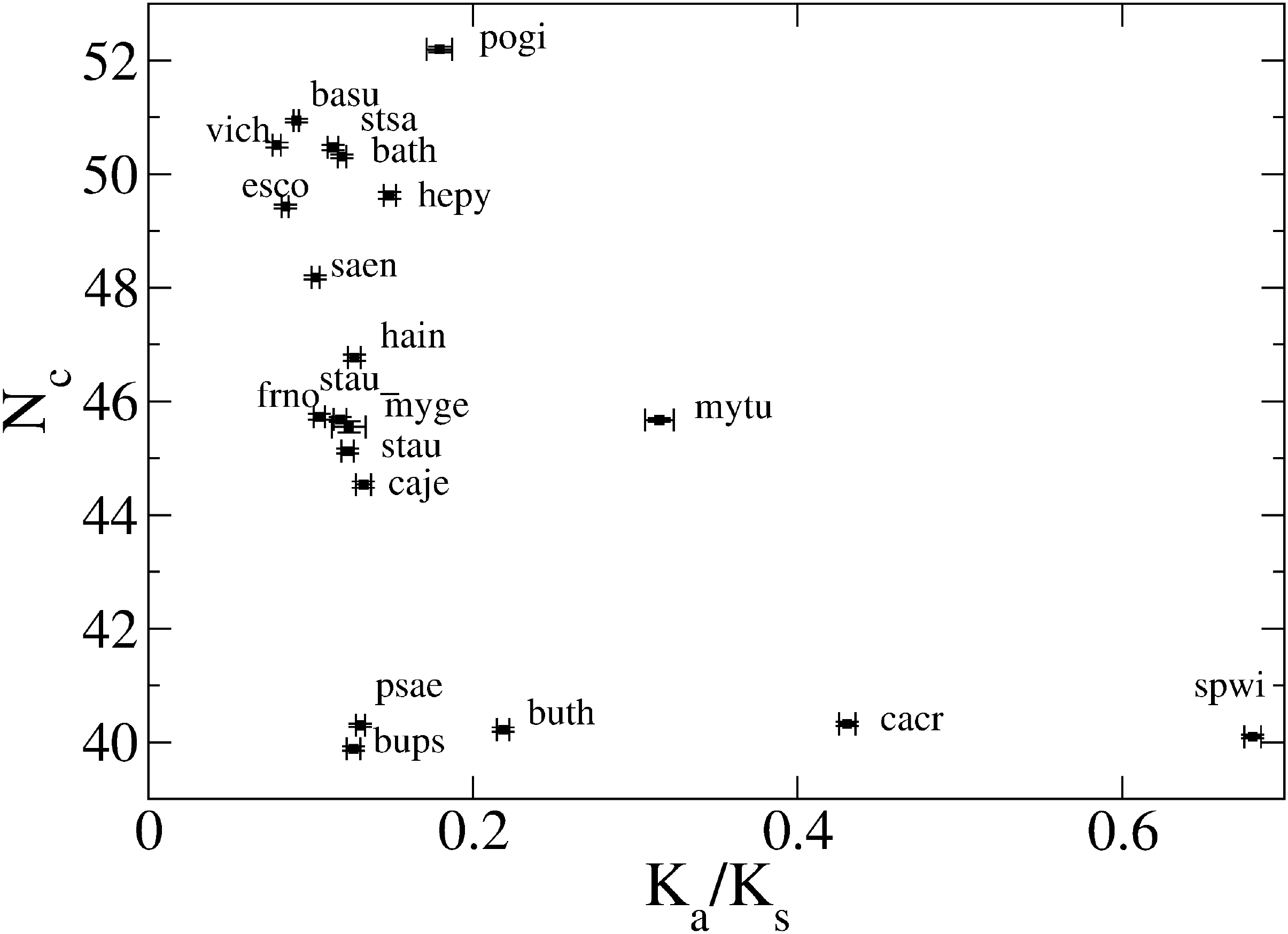}
\caption{Average values of (unnormalised) $N_c$ and $K_a/K_s$ for each bacterial species, with error bars denoting root mean square errors.}
\label{fig10}      
\end{figure}

As a conclusive observation, we correlate average $N_c$ values with corresponding average value 
of $K_a/K_s$ in different bacterial genomes (Figure~\ref{fig10}). 
Bacterial species appear to be separated in at least three clusters, 
corresponding to different ranges of average values of $N_c$, 
and average values of $K_a/K_s$ are consistent with the frequency distribution of Figure~\ref{fig7}. 
The few outliers, namely 11 ($buth$), 13 ($cacr$) and 19 ($spwi$), are the species 
with the highest $K_a/K_s$ ratios and the lowest $N_c$ values: 
an optimized choice of codons seems to be required to be under a more selective evolutive pressure, 
remember that lower values of $N_c$ indicate more selective choice of synonymous codons.
It would be interesting to have data on other bacterial genomes to complete the phase diagram 
correlating codon bias with evolutionary pressure, of which our Figure~\ref{fig10} 
is just a preliminary sketch, in order to deeply investigate 
the possibly subtle connection between codon bias at the genetic level 
with the propensity to mutate at the protein sequence level.

\begin{figure*}
 \includegraphics[width=8.6cm]{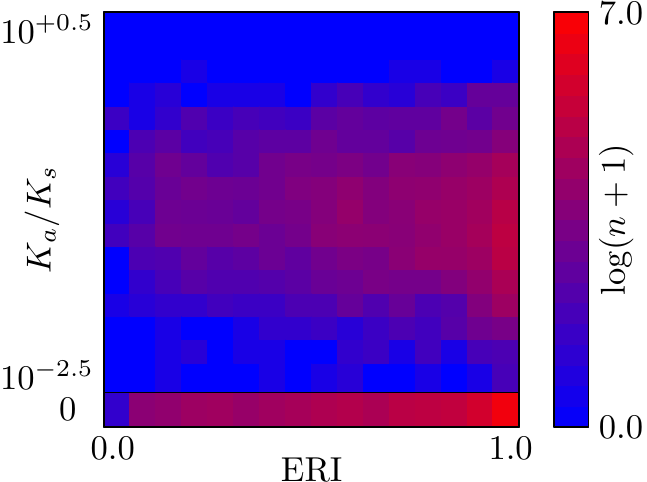}
 \includegraphics[width=8.6cm]{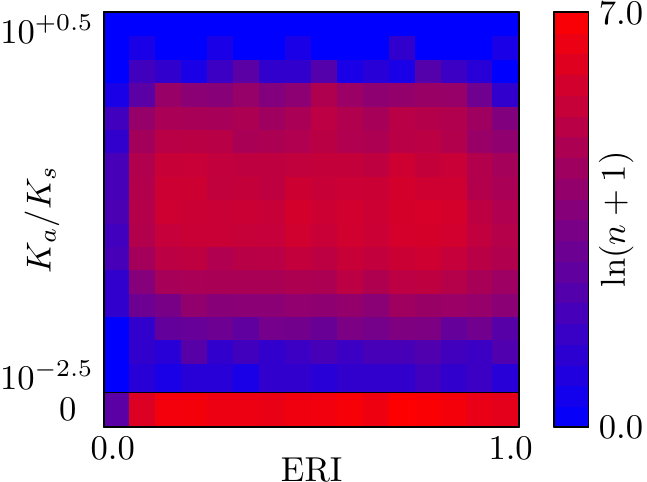}\\
 \includegraphics[width=8.6cm]{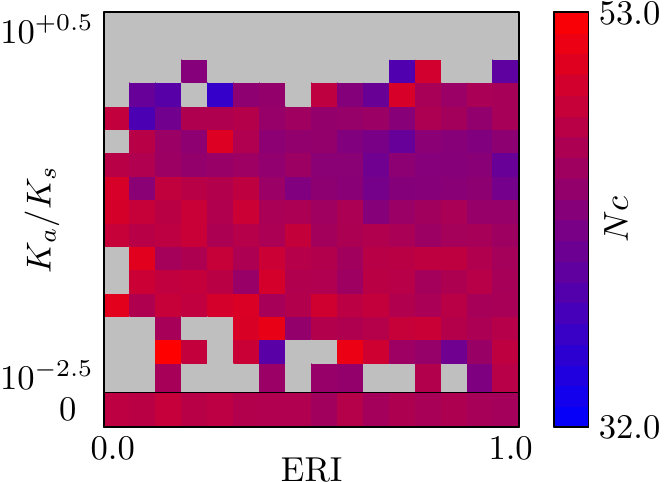}
 \includegraphics[width=8.6cm]{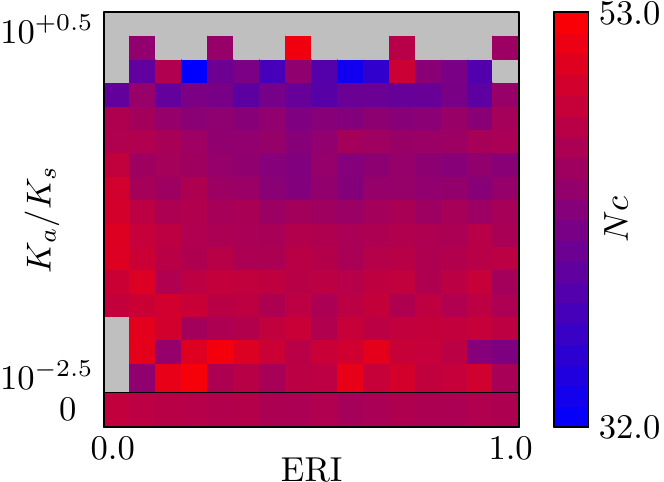}
\caption{Heat maps showing the number (upper panels) and average $N_c$ value (lower panels) 
for genes with given values of $ERI$ and $K_a/K_s$. We consider each genome in our database but split genes into essential and nonessential ones (left and right panels, respectively). 
Note that the lower (boxed) row of each panel corresponds to $K_a/K_s=0$, and that in the lower panels empty cells indicate absence of genes. 
Upper panels do show that the distribution of $K_a/K_s$ values is highly heterogeneous whatever the ERI values of genes. 
Lower panels show that $N_c$ slightly decreases with increasing ERI, but also seems to decrease for increasing $K_a/K_s$.}
\label{fig3d}      
\end{figure*}

\section*{Conclusions}

Inspired by the results by \citet{Luo2015}, with this work we further contribute
to the elucidation of the intricate connections among gene essentiality, conservation, 
codon usage bias and evolutionary pressure. In particular, we extended the investigation 
we performed on \EC~\citep{Dilucca2015} to several bacterial species. 
That essentiality, conservation, evolutionary pressure and codon bias in bacterial genes, and also in general should be strongly connected is
one of those views that are widely shared among life scientist. A view that rests on a broad literature. Our work does not convey radically new messages
and the results we have shown largely confirm shared views, but the merit of our observations resides in having shown, carefully, 
quantitative correlations that have been extended to several genomes. 
A unified view or theoretical model of the complex interplay among the quantities we have considered here is not yet available. 
We made a quite unsatisfactory attempt at a ternary representation of codon usage, retention index and selective pressure in the same plot, for essential and non essential genes. 
The heat maps we get in Figure \ref{fig3d} do not convey any clear emerging pattern and we believe that going beyond binary correlations 
would require in the future a focused effort by several researchers.  

Going back to our findings: as a first result, we have shown that there is a universal exponential 
correlation between gene essentiality and degree of conservation: genes with 
high values of the evolutionary retention index (ERI) are more likely to be essential (Figure~\ref{fig1}). 
We have then observed that the number of essential genes is rather conserved 
among bacterial species. Small bacterial genomes are composed mainly by 
essential genes but, as the size of the genome increases, the number of 
nonessential genes increases proportionally (Figure~\ref{fig2}). The set of 
around 500 essential genes in a given bacterial genome is however not composed by genes 
having orthologs in all the species: essentiality does not imply orthology. 
This is true also for the core of 83 genes which are strictly retained ($\mbox{ERI}=1$) 
in all the species here considered. These genes have a peculiar functional repertoire 
(mainly COGs J, but also K, L and O, see Table~\ref{tab.core}), and while 
they are not always essential they have, however, a probability of being essential not less than 0.56. 
This set could thus represent an optimal reservoir of potential targets 
for new antimicrobial components \citep{Fields2016}.

Regarding functional classification, we have considered how the 
different clusters of orthologous genes (COGs) accommodate essential and nonessential 
genes (Figure~\ref{fig3}). These two groups turn out to have a complementary spectrum of functions 
(Table~\ref{tab.func}): essential genes mainly fall into COGs J, M, H, and L, 
and prevail in functions related to information processing 
(translation/transcription, replication, recombination and repair), 
whereas, nonessential genes mainly belong to COGs E, K, G, C and P, 
with prevalence in metabolic functions (production and transport of energy and basic cellular constituents). 
Since essentiality implies a certain degree of evolutionary conservation, 
genes and functions of the first group of COGs should be under a relatively purifying selection, 
whereas, the second group of functions should be more prone to diversifying selection.
Indeed, we have shown in Figures~\ref{fig4} and~\ref{fig5} that more conserved (shared) genes 
feature significantly lower values of $K_a/K_s$ than less conserved genes. 

The distribution of $K_a/K_s$ values of Figure~\ref{fig7} shows that, overall, 
bacterial genes are under purifying evolutionary pressure, as $K_a/K_s$ 
is hardly greater than 1. Nevertheless, through the relative comparison 
of the individual Z-scores of $K_a/K_s$ for genes in different genomes and 
in different COGs of Figure~\ref{fig6}, we could sensibly assess that the different COGs 
are under different evolutionary pressure and constraints. We have thus proposed 
a new tentative distinction between a set of relatively more evolutionarily conserved COGs 
(J, F, K, and O) and a set of more adaptive ones (P, L, N, and M). 
Such a distinction is clearly at variance with the one coming from the analysis 
of essential genes we discussed above. Detailing the terms of this contradiction 
requires further investigation, particularly for understanding 
the relevance of the coverage of the databases for the consistency between the 
test based on the COG enrichment in essential genes with that based on the Z-scores of $K_a/K_s$.

Using $RSCU$ vectors and the effective number of codons $N_c$, we have finally shown that 
it is possible to finely classify bacteria following their codon usage patterns (Figure~\ref{fig8}). 
This classification still requires a consistent interpretation, possibly based 
on the analysis of ecological relationships among species. We have also shown in Figure~\ref{fig9} 
that specific and conserved (shared) genes make slightly different use of synonymous codons: 
more conserved genes have a reduced number of effective codons, a clear indication that 
conservation of a gene rests on some kind of evolutionary optimization in the use of synonymous codons. 
Distinguishing essential from nonessential genes does not change the overall classification, 
indicating that each bacterial species has its own strong signature in codon bias. 
This specificity of the bias suggest where to proceed in the next future, 
with further investigations on the relevance of codon bias in phylogeny reconstructions 
and in the prediction of protein-protein interaction networks.

\begin{center}
 * * *
\end{center}

We warmly thank Dr. Hao Luo from the Department of Physics of Tianjin University for sharing with us his evaluations 
of $K_a/K_s$ for the genomes in Table~\ref{tab.dataset}, that are also used by \citet{Luo2015}.


\begin{thebibliography}{31}%
\makeatletter
\providecommand \@ifxundefined [1]{%
 \@ifx{#1\undefined}
}%
\providecommand \@ifnum [1]{%
 \ifnum #1\expandafter \@firstoftwo
 \else \expandafter \@secondoftwo
 \fi
}%
\providecommand \@ifx [1]{%
 \ifx #1\expandafter \@firstoftwo
 \else \expandafter \@secondoftwo
 \fi
}%
\providecommand \natexlab [1]{#1}%
\providecommand \enquote  [1]{``#1''}%
\providecommand \bibnamefont  [1]{#1}%
\providecommand \bibfnamefont [1]{#1}%
\providecommand \citenamefont [1]{#1}%
\providecommand \href@noop [0]{\@secondoftwo}%
\providecommand \href [0]{\begingroup \@sanitize@url \@href}%
\providecommand \@href[1]{\@@startlink{#1}\@@href}%
\providecommand \@@href[1]{\endgroup#1\@@endlink}%
\providecommand \@sanitize@url [0]{\catcode `\\12\catcode `\$12\catcode
  `\&12\catcode `\#12\catcode `\^12\catcode `\_12\catcode `\%12\relax}%
\providecommand \@@startlink[1]{}%
\providecommand \@@endlink[0]{}%
\providecommand \url  [0]{\begingroup\@sanitize@url \@url }%
\providecommand \@url [1]{\endgroup\@href {#1}{\urlprefix }}%
\providecommand \urlprefix  [0]{URL }%
\providecommand \Eprint [0]{\href }%
\providecommand \doibase [0]{http://dx.doi.org/}%
\providecommand \selectlanguage [0]{\@gobble}%
\providecommand \bibinfo  [0]{\@secondoftwo}%
\providecommand \bibfield  [0]{\@secondoftwo}%
\providecommand \translation [1]{[#1]}%
\providecommand \BibitemOpen [0]{}%
\providecommand \bibitemStop [0]{}%
\providecommand \bibitemNoStop [0]{.\EOS\space}%
\providecommand \EOS [0]{\spacefactor3000\relax}%
\providecommand \BibitemShut  [1]{\csname bibitem#1\endcsname}%
\let\auto@bib@innerbib\@empty
\bibitem [{\citenamefont {Dilucca}\ \emph {et~al.}(2015)\citenamefont
  {Dilucca}, \citenamefont {Cimini}, \citenamefont {Semmoloni}, \citenamefont
  {Deiana},\ and\ \citenamefont {Giansanti}}]{Dilucca2015}%
  \BibitemOpen
  \bibfield  {author} {\bibinfo {author} {\bibfnamefont {M.}~\bibnamefont
  {Dilucca}}, \bibinfo {author} {\bibfnamefont {G.}~\bibnamefont {Cimini}},
  \bibinfo {author} {\bibfnamefont {A.}~\bibnamefont {Semmoloni}}, \bibinfo
  {author} {\bibfnamefont {A.}~\bibnamefont {Deiana}}, \ and\ \bibinfo {author}
  {\bibfnamefont {A.}~\bibnamefont {Giansanti}},\ }\href {\doibase
  10.1371/journal.pone.0142127} {\bibfield  {journal} {\bibinfo  {journal}
  {PLoS ONE}\ }\textbf {\bibinfo {volume} {10}},\ \bibinfo {pages} {e0142127}
  (\bibinfo {year} {2015})}\BibitemShut {NoStop}%
\bibitem [{\citenamefont {Gerdes}\ \emph {et~al.}(2003)\citenamefont {Gerdes},
  \citenamefont {Scholle}, \citenamefont {Campbell}, \citenamefont {Balazsi},
  \citenamefont {Ravasz}, \citenamefont {Daugherty}, \citenamefont {Somera},
  \citenamefont {Kyrpides}, \citenamefont {Anderson}, \citenamefont {Gelfand}
  \emph {et~al.}}]{Gerdes2003}%
  \BibitemOpen
  \bibfield  {author} {\bibinfo {author} {\bibfnamefont {S.}~\bibnamefont
  {Gerdes}}, \bibinfo {author} {\bibfnamefont {M.}~\bibnamefont {Scholle}},
  \bibinfo {author} {\bibfnamefont {J.}~\bibnamefont {Campbell}}, \bibinfo
  {author} {\bibfnamefont {G.}~\bibnamefont {Balazsi}}, \bibinfo {author}
  {\bibfnamefont {E.}~\bibnamefont {Ravasz}}, \bibinfo {author} {\bibfnamefont
  {M.}~\bibnamefont {Daugherty}}, \bibinfo {author} {\bibfnamefont
  {A.}~\bibnamefont {Somera}}, \bibinfo {author} {\bibfnamefont
  {N.}~\bibnamefont {Kyrpides}}, \bibinfo {author} {\bibfnamefont
  {I.}~\bibnamefont {Anderson}}, \bibinfo {author} {\bibfnamefont
  {M.}~\bibnamefont {Gelfand}},  \emph {et~al.},\ }\href {\doibase
  10.1128/JB.185.19.5673-5684.2003} {\bibfield  {journal} {\bibinfo  {journal}
  {Journal of Bacteriology}\ }\textbf {\bibinfo {volume} {185}},\ \bibinfo
  {pages} {5673} (\bibinfo {year} {2003})}\BibitemShut {NoStop}%
\bibitem [{\citenamefont {Fang}\ \emph {et~al.}(2005)\citenamefont {Fang},
  \citenamefont {Rocha},\ and\ \citenamefont {Danchin}}]{Fang2005}%
  \BibitemOpen
  \bibfield  {author} {\bibinfo {author} {\bibfnamefont {G.}~\bibnamefont
  {Fang}}, \bibinfo {author} {\bibfnamefont {E.}~\bibnamefont {Rocha}}, \ and\
  \bibinfo {author} {\bibfnamefont {A.}~\bibnamefont {Danchin}},\ }\href
  {\doibase 10.1093/molbev/msi211} {\bibfield  {journal} {\bibinfo  {journal}
  {Molecular Biology and Evolution}\ }\textbf {\bibinfo {volume} {22}},\
  \bibinfo {pages} {2147} (\bibinfo {year} {2005})}\BibitemShut {NoStop}%
\bibitem [{\citenamefont {Peng}\ and\ \citenamefont {Gao}(2014)}]{Peng2014}%
  \BibitemOpen
  \bibfield  {author} {\bibinfo {author} {\bibfnamefont {C.}~\bibnamefont
  {Peng}}\ and\ \bibinfo {author} {\bibfnamefont {F.}~\bibnamefont {Gao}},\
  }\href {\doibase 10.1038/srep06001} {\bibfield  {journal} {\bibinfo
  {journal} {Scientific Reports}\ }\textbf {\bibinfo {volume} {4}},\ \bibinfo
  {pages} {6001} (\bibinfo {year} {2014})}\BibitemShut {NoStop}%
\bibitem [{\citenamefont {Lin}\ \emph {et~al.}(2010)\citenamefont {Lin},
  \citenamefont {Gao},\ and\ \citenamefont {Zhang}}]{Lin2010}%
  \BibitemOpen
  \bibfield  {author} {\bibinfo {author} {\bibfnamefont {Y.}~\bibnamefont
  {Lin}}, \bibinfo {author} {\bibfnamefont {F.}~\bibnamefont {Gao}}, \ and\
  \bibinfo {author} {\bibfnamefont {C.-T.}\ \bibnamefont {Zhang}},\ }\href
  {\doibase 10.1016/j.bbrc.2010.04.119} {\bibfield  {journal} {\bibinfo
  {journal} {Biochemical and Biophysical Research Communications}\ }\textbf
  {\bibinfo {volume} {396}},\ \bibinfo {pages} {472} (\bibinfo {year}
  {2010})}\BibitemShut {NoStop}%
\bibitem [{\citenamefont {Hurst}\ and\ \citenamefont
  {Smith}(1999)}]{Hurst1999}%
  \BibitemOpen
  \bibfield  {author} {\bibinfo {author} {\bibfnamefont {L.~D.}\ \bibnamefont
  {Hurst}}\ and\ \bibinfo {author} {\bibfnamefont {N.~G.}\ \bibnamefont
  {Smith}},\ }\href {\doibase 10.1016/S0960-9822(99)80334-0} {\bibfield
  {journal} {\bibinfo  {journal} {Current Biology}\ }\textbf {\bibinfo {volume}
  {9}},\ \bibinfo {pages} {747} (\bibinfo {year} {1999})}\BibitemShut {NoStop}%
\bibitem [{\citenamefont {Jordan}\ \emph {et~al.}(2002)\citenamefont {Jordan},
  \citenamefont {Rogozin}, \citenamefont {Wolf},\ and\ \citenamefont
  {Koonin}}]{Jordan2002}%
  \BibitemOpen
  \bibfield  {author} {\bibinfo {author} {\bibfnamefont {I.~K.}\ \bibnamefont
  {Jordan}}, \bibinfo {author} {\bibfnamefont {I.~B.}\ \bibnamefont {Rogozin}},
  \bibinfo {author} {\bibfnamefont {Y.~I.}\ \bibnamefont {Wolf}}, \ and\
  \bibinfo {author} {\bibfnamefont {E.~V.}\ \bibnamefont {Koonin}},\ }\href
  {\doibase 10.1101/gr.87702} {\bibfield  {journal} {\bibinfo  {journal}
  {Genome research}\ }\textbf {\bibinfo {volume} {12}},\ \bibinfo {pages} {962}
  (\bibinfo {year} {2002})}\BibitemShut {NoStop}%
\bibitem [{\citenamefont {Luo}\ \emph {et~al.}(2015)\citenamefont {Luo},
  \citenamefont {Gao},\ and\ \citenamefont {Lin}}]{Luo2015}%
  \BibitemOpen
  \bibfield  {author} {\bibinfo {author} {\bibfnamefont {H.}~\bibnamefont
  {Luo}}, \bibinfo {author} {\bibfnamefont {F.}~\bibnamefont {Gao}}, \ and\
  \bibinfo {author} {\bibfnamefont {Y.}~\bibnamefont {Lin}},\ }\href {\doibase
  10.1038/srep13210} {\bibfield  {journal} {\bibinfo  {journal} {Scientific
  Reports}\ }\textbf {\bibinfo {volume} {5}},\ \bibinfo {pages} {13210}
  (\bibinfo {year} {2015})}\BibitemShut {NoStop}%
\bibitem [{\citenamefont {Ish-Am}\ \emph {et~al.}(2015)\citenamefont {Ish-Am},
  \citenamefont {Kristensen},\ and\ \citenamefont {Ruppin}}]{Ish-Am2015}%
  \BibitemOpen
  \bibfield  {author} {\bibinfo {author} {\bibfnamefont {O.}~\bibnamefont
  {Ish-Am}}, \bibinfo {author} {\bibfnamefont {D.~M.}\ \bibnamefont
  {Kristensen}}, \ and\ \bibinfo {author} {\bibfnamefont {E.}~\bibnamefont
  {Ruppin}},\ }\href {\doibase 10.1371/journal.pone.0123785} {\bibfield
  {journal} {\bibinfo  {journal} {PLoS ONE}\ }\textbf {\bibinfo {volume}
  {10}},\ \bibinfo {pages} {e0123785} (\bibinfo {year} {2015})}\BibitemShut
  {NoStop}%
\bibitem [{\citenamefont {Alvarez-Ponce}\ \emph {et~al.}(2016)\citenamefont
  {Alvarez-Ponce}, \citenamefont {Sabater-Mu{\~n}oz}, \citenamefont {Toft},
  \citenamefont {Ruiz-Gonz{\'a}lez},\ and\ \citenamefont
  {Fares}}]{Alvarez-Ponce2016}%
  \BibitemOpen
  \bibfield  {author} {\bibinfo {author} {\bibfnamefont {D.}~\bibnamefont
  {Alvarez-Ponce}}, \bibinfo {author} {\bibfnamefont {B.}~\bibnamefont
  {Sabater-Mu{\~n}oz}}, \bibinfo {author} {\bibfnamefont {C.}~\bibnamefont
  {Toft}}, \bibinfo {author} {\bibfnamefont {M.~X.}\ \bibnamefont
  {Ruiz-Gonz{\'a}lez}}, \ and\ \bibinfo {author} {\bibfnamefont {M.~A.}\
  \bibnamefont {Fares}},\ }\href {\doibase 10.1093/gbe/evw205} {\bibfield
  {journal} {\bibinfo  {journal} {Genome Biology and Evolution}\ }\textbf
  {\bibinfo {volume} {8}},\ \bibinfo {pages} {2914} (\bibinfo {year}
  {2016})}\BibitemShut {NoStop}%
\bibitem [{\citenamefont {Bergmiller}\ \emph {et~al.}(2012)\citenamefont
  {Bergmiller}, \citenamefont {Ackermann},\ and\ \citenamefont
  {Silander}}]{Bergmiller2012}%
  \BibitemOpen
  \bibfield  {author} {\bibinfo {author} {\bibfnamefont {T.}~\bibnamefont
  {Bergmiller}}, \bibinfo {author} {\bibfnamefont {M.}~\bibnamefont
  {Ackermann}}, \ and\ \bibinfo {author} {\bibfnamefont {O.~K.}\ \bibnamefont
  {Silander}},\ }\href {\doibase 10.1371/journal.pgen.1002803} {\bibfield
  {journal} {\bibinfo  {journal} {PLoS Genetics}\ }\textbf {\bibinfo {volume}
  {8}},\ \bibinfo {pages} {e1002803} (\bibinfo {year} {2012})}\BibitemShut
  {NoStop}%
\bibitem [{\citenamefont {Hurst}(2002)}]{Hurst2002}%
  \BibitemOpen
  \bibfield  {author} {\bibinfo {author} {\bibfnamefont {L.~D.}\ \bibnamefont
  {Hurst}},\ }\href {\doibase 10.1016/S0168-9525(02)02722-1} {\bibfield
  {journal} {\bibinfo  {journal} {Trends in Genetics}\ }\textbf {\bibinfo
  {volume} {18}},\ \bibinfo {pages} {486} (\bibinfo {year} {2002})}\BibitemShut
  {NoStop}%
\bibitem [{\citenamefont {Kanaya}\ \emph {et~al.}(2001)\citenamefont {Kanaya},
  \citenamefont {Yamada}, \citenamefont {Kinouchi}, \citenamefont {Kudo},\ and\
  \citenamefont {Ikemura}}]{Kanaya2001}%
  \BibitemOpen
  \bibfield  {author} {\bibinfo {author} {\bibfnamefont {S.}~\bibnamefont
  {Kanaya}}, \bibinfo {author} {\bibfnamefont {Y.}~\bibnamefont {Yamada}},
  \bibinfo {author} {\bibfnamefont {M.}~\bibnamefont {Kinouchi}}, \bibinfo
  {author} {\bibfnamefont {Y.}~\bibnamefont {Kudo}}, \ and\ \bibinfo {author}
  {\bibfnamefont {T.}~\bibnamefont {Ikemura}},\ }\href {\doibase
  10.1007/s002390010219} {\bibfield  {journal} {\bibinfo  {journal} {Journal of
  Molecular Evolution}\ }\textbf {\bibinfo {volume} {53}},\ \bibinfo {pages}
  {290} (\bibinfo {year} {2001})}\BibitemShut {NoStop}%
\bibitem [{\citenamefont {Roth}\ \emph {et~al.}(2012)\citenamefont {Roth},
  \citenamefont {Anisimova},\ and\ \citenamefont {Cannarozzi}}]{Roth2012}%
  \BibitemOpen
  \bibfield  {author} {\bibinfo {author} {\bibfnamefont {A.}~\bibnamefont
  {Roth}}, \bibinfo {author} {\bibfnamefont {M.}~\bibnamefont {Anisimova}}, \
  and\ \bibinfo {author} {\bibfnamefont {G.~M.}\ \bibnamefont {Cannarozzi}},\
  }\enquote {\bibinfo {title} {Measuring codon usage bias},}\ in\ \href@noop {}
  {\emph {\bibinfo {booktitle} {Codon evolution: mechanisms and models}}}\
  (\bibinfo  {publisher} {Oxford University Press Inc. (New York)},\ \bibinfo
  {year} {2012})\ pp.\ \bibinfo {pages} {189--217}\BibitemShut {NoStop}%
\bibitem [{\citenamefont {D{\"o}tsch}\ \emph {et~al.}(2010)\citenamefont
  {D{\"o}tsch}, \citenamefont {Klawonn}, \citenamefont {Jarek}, \citenamefont
  {Scharfe}, \citenamefont {Bl{\"o}cker},\ and\ \citenamefont
  {H{\"a}ussler}}]{Dotsch2010}%
  \BibitemOpen
  \bibfield  {author} {\bibinfo {author} {\bibfnamefont {A.}~\bibnamefont
  {D{\"o}tsch}}, \bibinfo {author} {\bibfnamefont {F.}~\bibnamefont {Klawonn}},
  \bibinfo {author} {\bibfnamefont {M.}~\bibnamefont {Jarek}}, \bibinfo
  {author} {\bibfnamefont {M.}~\bibnamefont {Scharfe}}, \bibinfo {author}
  {\bibfnamefont {H.}~\bibnamefont {Bl{\"o}cker}}, \ and\ \bibinfo {author}
  {\bibfnamefont {S.}~\bibnamefont {H{\"a}ussler}},\ }\href {\doibase
  10.1186/1471-2164-11-234} {\bibfield  {journal} {\bibinfo  {journal} {BMC
  genomics}\ }\textbf {\bibinfo {volume} {11}},\ \bibinfo {pages} {234}
  (\bibinfo {year} {2010})}\BibitemShut {NoStop}%
\bibitem [{\citenamefont {Benson}\ \emph {et~al.}(2013)\citenamefont {Benson},
  \citenamefont {Cavanaugh}, \citenamefont {Clark}, \citenamefont
  {Karsch-Mizrachi}, \citenamefont {Lipman}, \citenamefont {Ostell},\ and\
  \citenamefont {Sayers}}]{Benson2012}%
  \BibitemOpen
  \bibfield  {author} {\bibinfo {author} {\bibfnamefont {D.~A.}\ \bibnamefont
  {Benson}}, \bibinfo {author} {\bibfnamefont {M.}~\bibnamefont {Cavanaugh}},
  \bibinfo {author} {\bibfnamefont {K.}~\bibnamefont {Clark}}, \bibinfo
  {author} {\bibfnamefont {I.}~\bibnamefont {Karsch-Mizrachi}}, \bibinfo
  {author} {\bibfnamefont {D.~J.}\ \bibnamefont {Lipman}}, \bibinfo {author}
  {\bibfnamefont {J.}~\bibnamefont {Ostell}}, \ and\ \bibinfo {author}
  {\bibfnamefont {E.~W.}\ \bibnamefont {Sayers}},\ }\href {\doibase
  10.1093/nar/gks1195} {\bibfield  {journal} {\bibinfo  {journal} {Nucleic
  Acids Research}\ }\textbf {\bibinfo {volume} {41}},\ \bibinfo {pages} {D36}
  (\bibinfo {year} {2013})}\BibitemShut {NoStop}%
\bibitem [{\citenamefont {Zhang}\ and\ \citenamefont {Lin}(2009)}]{Zhang2009}%
  \BibitemOpen
  \bibfield  {author} {\bibinfo {author} {\bibfnamefont {R.}~\bibnamefont
  {Zhang}}\ and\ \bibinfo {author} {\bibfnamefont {Y.}~\bibnamefont {Lin}},\
  }\href@noop {} {\bibfield  {journal} {\bibinfo  {journal} {Nucleic Acids
  Research}\ }\textbf {\bibinfo {volume} {37}},\ \bibinfo {pages} {D455}
  (\bibinfo {year} {2009})}\BibitemShut {NoStop}%
\bibitem [{\citenamefont {Luo}\ \emph {et~al.}(2014)\citenamefont {Luo},
  \citenamefont {Lin}, \citenamefont {Gao}, \citenamefont {Zhang},\ and\
  \citenamefont {Zhang}}]{Zhang2014}%
  \BibitemOpen
  \bibfield  {author} {\bibinfo {author} {\bibfnamefont {H.}~\bibnamefont
  {Luo}}, \bibinfo {author} {\bibfnamefont {Y.}~\bibnamefont {Lin}}, \bibinfo
  {author} {\bibfnamefont {F.}~\bibnamefont {Gao}}, \bibinfo {author}
  {\bibfnamefont {C.-T.}\ \bibnamefont {Zhang}}, \ and\ \bibinfo {author}
  {\bibfnamefont {R.}~\bibnamefont {Zhang}},\ }\href {\doibase
  10.1093/nar/gkt1131} {\bibfield  {journal} {\bibinfo  {journal} {Nucleic
  Acids Research}\ }\textbf {\bibinfo {volume} {42}},\ \bibinfo {pages} {D574}
  (\bibinfo {year} {2014})}\BibitemShut {NoStop}%
\bibitem [{\citenamefont {Galperin}\ \emph {et~al.}(2015)\citenamefont
  {Galperin}, \citenamefont {Makarova}, \citenamefont {Wolf},\ and\
  \citenamefont {Koonin}}]{Galperin2015}%
  \BibitemOpen
  \bibfield  {author} {\bibinfo {author} {\bibfnamefont {M.~Y.}\ \bibnamefont
  {Galperin}}, \bibinfo {author} {\bibfnamefont {K.~S.}\ \bibnamefont
  {Makarova}}, \bibinfo {author} {\bibfnamefont {Y.~I.}\ \bibnamefont {Wolf}},
  \ and\ \bibinfo {author} {\bibfnamefont {E.~V.}\ \bibnamefont {Koonin}},\
  }\href {\doibase 10.1093/nar/gku1223} {\bibfield  {journal} {\bibinfo
  {journal} {Nucleic Acids Research}\ }\textbf {\bibinfo {volume} {43}},\
  \bibinfo {pages} {D261} (\bibinfo {year} {2015})}\BibitemShut {NoStop}%
\bibitem [{\citenamefont {Nei}\ and\ \citenamefont {Gojobori}(1986)}]{Nei1986}%
  \BibitemOpen
  \bibfield  {author} {\bibinfo {author} {\bibfnamefont {M.}~\bibnamefont
  {Nei}}\ and\ \bibinfo {author} {\bibfnamefont {T.}~\bibnamefont {Gojobori}},\
  }\href {\doibase 10.1093/oxfordjournals.molbev.a040410} {\bibfield  {journal}
  {\bibinfo  {journal} {Molecular Biology and Evolution}\ }\textbf {\bibinfo
  {volume} {3}},\ \bibinfo {pages} {418} (\bibinfo {year} {1986})}\BibitemShut
  {NoStop}%
\bibitem [{\citenamefont {Wright}(1990)}]{Wright1990}%
  \BibitemOpen
  \bibfield  {author} {\bibinfo {author} {\bibfnamefont {F.}~\bibnamefont
  {Wright}},\ }\href {\doibase 10.1016/0378-1119(90)90491-9} {\bibfield
  {journal} {\bibinfo  {journal} {Gene}\ }\textbf {\bibinfo {volume} {87}},\
  \bibinfo {pages} {23} (\bibinfo {year} {1990})}\BibitemShut {NoStop}%
\bibitem [{\citenamefont {Xia}(2013)}]{Xia2013}%
  \BibitemOpen
  \bibfield  {author} {\bibinfo {author} {\bibfnamefont {X.}~\bibnamefont
  {Xia}},\ }\href {\doibase 10.1093/molbev/mst064} {\bibfield  {journal}
  {\bibinfo  {journal} {Molecular Biology and Evolution}\ }\textbf {\bibinfo
  {volume} {30}},\ \bibinfo {pages} {1720} (\bibinfo {year}
  {2013})}\BibitemShut {NoStop}%
\bibitem [{\citenamefont {Gong}\ \emph {et~al.}(2008)\citenamefont {Gong},
  \citenamefont {Fan}, \citenamefont {Bilderbeck}, \citenamefont {Li},
  \citenamefont {Pang},\ and\ \citenamefont {Tao}}]{Gong2008}%
  \BibitemOpen
  \bibfield  {author} {\bibinfo {author} {\bibfnamefont {X.}~\bibnamefont
  {Gong}}, \bibinfo {author} {\bibfnamefont {S.}~\bibnamefont {Fan}}, \bibinfo
  {author} {\bibfnamefont {A.}~\bibnamefont {Bilderbeck}}, \bibinfo {author}
  {\bibfnamefont {M.}~\bibnamefont {Li}}, \bibinfo {author} {\bibfnamefont
  {H.}~\bibnamefont {Pang}}, \ and\ \bibinfo {author} {\bibfnamefont
  {S.}~\bibnamefont {Tao}},\ }\href {\doibase 10.1007/s00438-007-0298-x}
  {\bibfield  {journal} {\bibinfo  {journal} {Molecular Genetics and Genomics}\
  }\textbf {\bibinfo {volume} {279}},\ \bibinfo {pages} {87} (\bibinfo {year}
  {2008})}\BibitemShut {NoStop}%
\bibitem [{\citenamefont {Gil}\ \emph {et~al.}(2004)\citenamefont {Gil},
  \citenamefont {Silva}, \citenamefont {Peret{\'o}},\ and\ \citenamefont
  {Moya}}]{Gil2004}%
  \BibitemOpen
  \bibfield  {author} {\bibinfo {author} {\bibfnamefont {R.}~\bibnamefont
  {Gil}}, \bibinfo {author} {\bibfnamefont {F.~J.}\ \bibnamefont {Silva}},
  \bibinfo {author} {\bibfnamefont {J.}~\bibnamefont {Peret{\'o}}}, \ and\
  \bibinfo {author} {\bibfnamefont {A.}~\bibnamefont {Moya}},\ }\href {\doibase
  10.1128/MMBR.68.3.518-537.2004} {\bibfield  {journal} {\bibinfo  {journal}
  {Microbiology and Molecular Biology Reviews}\ }\textbf {\bibinfo {volume}
  {68}},\ \bibinfo {pages} {518} (\bibinfo {year} {2004})}\BibitemShut
  {NoStop}%
\bibitem [{\citenamefont {Glass}\ \emph {et~al.}(2006)\citenamefont {Glass},
  \citenamefont {Assad-Garcia}, \citenamefont {Alperovich}, \citenamefont
  {Yooseph}, \citenamefont {Lewis}, \citenamefont {Maruf}, \citenamefont
  {Hutchison}, \citenamefont {Smith},\ and\ \citenamefont
  {Venter}}]{Glass2006}%
  \BibitemOpen
  \bibfield  {author} {\bibinfo {author} {\bibfnamefont {J.~I.}\ \bibnamefont
  {Glass}}, \bibinfo {author} {\bibfnamefont {N.}~\bibnamefont {Assad-Garcia}},
  \bibinfo {author} {\bibfnamefont {N.}~\bibnamefont {Alperovich}}, \bibinfo
  {author} {\bibfnamefont {S.}~\bibnamefont {Yooseph}}, \bibinfo {author}
  {\bibfnamefont {M.~R.}\ \bibnamefont {Lewis}}, \bibinfo {author}
  {\bibfnamefont {M.}~\bibnamefont {Maruf}}, \bibinfo {author} {\bibfnamefont
  {C.~A.}\ \bibnamefont {Hutchison}}, \bibinfo {author} {\bibfnamefont {H.~O.}\
  \bibnamefont {Smith}}, \ and\ \bibinfo {author} {\bibfnamefont {J.~C.}\
  \bibnamefont {Venter}},\ }\href {\doibase 10.1073/pnas.0510013103} {\bibfield
   {journal} {\bibinfo  {journal} {Proceedings of the National Academy of
  Sciences}\ }\textbf {\bibinfo {volume} {103}},\ \bibinfo {pages} {425}
  (\bibinfo {year} {2006})}\BibitemShut {NoStop}%
\bibitem [{\citenamefont {Hutchison}\ \emph {et~al.}(2016)\citenamefont
  {Hutchison}, \citenamefont {Chuang}, \citenamefont {Noskov}, \citenamefont
  {Assad-Garcia}, \citenamefont {Deerinck}, \citenamefont {Ellisman},
  \citenamefont {Gill}, \citenamefont {Kannan}, \citenamefont {Karas},
  \citenamefont {Ma}, \citenamefont {Pelletier}, \citenamefont {Qi},
  \citenamefont {Richter}, \citenamefont {Strychalski}, \citenamefont {Sun},
  \citenamefont {Suzuki}, \citenamefont {Tsvetanova}, \citenamefont {Wise},
  \citenamefont {Smith}, \citenamefont {Glass}, \citenamefont {Merryman},
  \citenamefont {Gibson},\ and\ \citenamefont {Venter}}]{Hutchison2016}%
  \BibitemOpen
  \bibfield  {author} {\bibinfo {author} {\bibfnamefont {C.~A.}\ \bibnamefont
  {Hutchison}}, \bibinfo {author} {\bibfnamefont {R.-Y.}\ \bibnamefont
  {Chuang}}, \bibinfo {author} {\bibfnamefont {V.~N.}\ \bibnamefont {Noskov}},
  \bibinfo {author} {\bibfnamefont {N.}~\bibnamefont {Assad-Garcia}}, \bibinfo
  {author} {\bibfnamefont {T.~J.}\ \bibnamefont {Deerinck}}, \bibinfo {author}
  {\bibfnamefont {M.~H.}\ \bibnamefont {Ellisman}}, \bibinfo {author}
  {\bibfnamefont {J.}~\bibnamefont {Gill}}, \bibinfo {author} {\bibfnamefont
  {K.}~\bibnamefont {Kannan}}, \bibinfo {author} {\bibfnamefont {B.~J.}\
  \bibnamefont {Karas}}, \bibinfo {author} {\bibfnamefont {L.}~\bibnamefont
  {Ma}}, \bibinfo {author} {\bibfnamefont {J.~F.}\ \bibnamefont {Pelletier}},
  \bibinfo {author} {\bibfnamefont {Z.-Q.}\ \bibnamefont {Qi}}, \bibinfo
  {author} {\bibfnamefont {R.~A.}\ \bibnamefont {Richter}}, \bibinfo {author}
  {\bibfnamefont {E.~A.}\ \bibnamefont {Strychalski}}, \bibinfo {author}
  {\bibfnamefont {L.}~\bibnamefont {Sun}}, \bibinfo {author} {\bibfnamefont
  {Y.}~\bibnamefont {Suzuki}}, \bibinfo {author} {\bibfnamefont
  {B.}~\bibnamefont {Tsvetanova}}, \bibinfo {author} {\bibfnamefont {K.~S.}\
  \bibnamefont {Wise}}, \bibinfo {author} {\bibfnamefont {H.~O.}\ \bibnamefont
  {Smith}}, \bibinfo {author} {\bibfnamefont {J.~I.}\ \bibnamefont {Glass}},
  \bibinfo {author} {\bibfnamefont {C.}~\bibnamefont {Merryman}}, \bibinfo
  {author} {\bibfnamefont {D.~G.}\ \bibnamefont {Gibson}}, \ and\ \bibinfo
  {author} {\bibfnamefont {J.~C.}\ \bibnamefont {Venter}},\ }\href {\doibase
  10.1126/science.aad6253} {\bibfield  {journal} {\bibinfo  {journal}
  {Science}\ }\textbf {\bibinfo {volume} {351}},\ \bibinfo {pages} {aad6253}
  (\bibinfo {year} {2016})}\BibitemShut {NoStop}%
\bibitem [{\citenamefont {Chessher}(2012)}]{Chessher2012}%
  \BibitemOpen
  \bibfield  {author} {\bibinfo {author} {\bibfnamefont {A.}~\bibnamefont
  {Chessher}},\ }\href {\doibase 10.1007/s13238-011-1135-x} {\bibfield
  {journal} {\bibinfo  {journal} {Protein {\&} Cell}\ }\textbf {\bibinfo
  {volume} {3}},\ \bibinfo {pages} {5} (\bibinfo {year} {2012})}\BibitemShut
  {NoStop}%
\bibitem [{\citenamefont {Koo}\ \emph {et~al.}(2017)\citenamefont {Koo},
  \citenamefont {Kritikos}, \citenamefont {Farelli}, \citenamefont {Todor},
  \citenamefont {Tong}, \citenamefont {Kimsey}, \citenamefont {Wapinski},
  \citenamefont {Galardini}, \citenamefont {Cabal}, \citenamefont {Peters},
  \citenamefont {Hachmann}, \citenamefont {Rudner}, \citenamefont {Allen},
  \citenamefont {Typas},\ and\ \citenamefont {Gross}}]{Koo2017}%
  \BibitemOpen
  \bibfield  {author} {\bibinfo {author} {\bibfnamefont {B.-M.}\ \bibnamefont
  {Koo}}, \bibinfo {author} {\bibfnamefont {G.}~\bibnamefont {Kritikos}},
  \bibinfo {author} {\bibfnamefont {J.~D.}\ \bibnamefont {Farelli}}, \bibinfo
  {author} {\bibfnamefont {H.}~\bibnamefont {Todor}}, \bibinfo {author}
  {\bibfnamefont {K.}~\bibnamefont {Tong}}, \bibinfo {author} {\bibfnamefont
  {H.}~\bibnamefont {Kimsey}}, \bibinfo {author} {\bibfnamefont
  {I.}~\bibnamefont {Wapinski}}, \bibinfo {author} {\bibfnamefont
  {M.}~\bibnamefont {Galardini}}, \bibinfo {author} {\bibfnamefont
  {A.}~\bibnamefont {Cabal}}, \bibinfo {author} {\bibfnamefont {J.~M.}\
  \bibnamefont {Peters}}, \bibinfo {author} {\bibfnamefont {A.-B.}\
  \bibnamefont {Hachmann}}, \bibinfo {author} {\bibfnamefont {D.~Z.}\
  \bibnamefont {Rudner}}, \bibinfo {author} {\bibfnamefont {K.~N.}\
  \bibnamefont {Allen}}, \bibinfo {author} {\bibfnamefont {A.}~\bibnamefont
  {Typas}}, \ and\ \bibinfo {author} {\bibfnamefont {C.~A.}\ \bibnamefont
  {Gross}},\ }\href {\doibase 10.1016/j.cels.2016.12.013} {\bibfield  {journal}
  {\bibinfo  {journal} {Cell Systems}\ }\textbf {\bibinfo {volume} {4}},\
  \bibinfo {pages} {291} (\bibinfo {year} {2017})}\BibitemShut {NoStop}%
\bibitem [{\citenamefont {Plotkin}\ and\ \citenamefont
  {Kudla}(2011)}]{Plotkin2011}%
  \BibitemOpen
  \bibfield  {author} {\bibinfo {author} {\bibfnamefont {J.~B.}\ \bibnamefont
  {Plotkin}}\ and\ \bibinfo {author} {\bibfnamefont {G.}~\bibnamefont
  {Kudla}},\ }\href {\doibase 10.1038/nrg2899} {\bibfield  {journal} {\bibinfo
  {journal} {Nature Reviews Genetics}\ }\textbf {\bibinfo {volume} {12}},\
  \bibinfo {pages} {32} (\bibinfo {year} {2011})}\BibitemShut {NoStop}%
\bibitem [{\citenamefont {Ran}\ and\ \citenamefont {Higgs}(2012)}]{Ran2012}%
  \BibitemOpen
  \bibfield  {author} {\bibinfo {author} {\bibfnamefont {W.}~\bibnamefont
  {Ran}}\ and\ \bibinfo {author} {\bibfnamefont {P.~G.}\ \bibnamefont
  {Higgs}},\ }\href {\doibase 10.1371/journal.pone.0051652} {\bibfield
  {journal} {\bibinfo  {journal} {PLoS ONE}\ }\textbf {\bibinfo {volume} {7}},\
  \bibinfo {pages} {e51652} (\bibinfo {year} {2012})}\BibitemShut {NoStop}%
\bibitem [{\citenamefont {Fields}\ \emph {et~al.}(2017)\citenamefont {Fields},
  \citenamefont {Lee},\ and\ \citenamefont {McConnell}}]{Fields2016}%
  \BibitemOpen
  \bibfield  {author} {\bibinfo {author} {\bibfnamefont {F.~R.}\ \bibnamefont
  {Fields}}, \bibinfo {author} {\bibfnamefont {S.~W.}\ \bibnamefont {Lee}}, \
  and\ \bibinfo {author} {\bibfnamefont {M.~J.}\ \bibnamefont {McConnell}},\
  }\href {\doibase 10.1016/j.bcp.2016.12.002} {\bibfield  {journal} {\bibinfo
  {journal} {Biochemical Pharmacology}\ }\textbf {\bibinfo {volume} {134}},\
  \bibinfo {pages} {74} (\bibinfo {year} {2017})}\BibitemShut {NoStop}%
\end{thebibliography}
%

\end{document}